\documentclass[equations,12pt]{article}
\usepackage{amsfonts}
\usepackage{a4,tikz}
\usepackage{geometry}
\usepackage{color} 
\usepackage{subfigure}
\makeatletter
\newcommand\ackname{Acknowledgements}
\if@titlepage
  \newenvironment{acknowledgements}{%
      \titlepage
      \null\vfil
      \@beginparpenalty\@lowpenalty
      \begin{center}%
        \bfseries \ackname
        \@endparpenalty\@M1
      \end{center}}%
     {\par\vfil\null\endtitlepage}
\else
  \newenvironment{acknowledgements}{%
      \if@twocolumn
        \section*{\abstractname}%
      \else
        \small
        \begin{center}%
          {\bfseries \ackname\vspace{-.5em}\vspace{\z@}}%
        \end{center}%
        \quotation
      \fi}
      {\if@twocolumn\else\endquotation\fi}
\fi
\makeatother
\usepackage{amsmath}
\usepackage{amssymb}
\usepackage{amsthm}
\usepackage{mathrsfs}
\usepackage{eucal}
 \usepackage[usenames,dvipsnames]{pstricks}
 \usepackage{epsfig}
 \usepackage{pst-grad} 
 \usepackage{pst-plot} 
\renewcommand{\theequation}{\arabic{equation}}
\usepackage{amssymb,amsmath}
\usepackage{color}
\usepackage{accents}
\usepackage{texdraw}
\usepackage{eucal}
\usepackage{epic,epsfig}
\usepackage{graphicx}
\usepackage{amsmath}
\usepackage{amssymb}
\usepackage{graphicx}

\theoremstyle{definition}

\numberwithin{equation}{section}
\DeclareMathAccent{\wtilde}{\mathord}{largesymbols}{"65}
\DeclareMathAccent{\what}{\mathord}{largesymbols}{"62}
\DeclareMathOperator{\csch}{csch}
\def\m@th{\mathsurround=0pt}
\mathchardef\bracell="0365
\def\upbrall{$\m@th\bracell$}
\def\undertilde#1{\mathop{\vtop{\ialign{##\crcr
    $\hfil\displaystyle{#1}\hfil$\crcr
     \noalign
     {\kern1.5pt\nointerlineskip}
     \upbrall\crcr\noalign{\kern1pt
   }}}}\limits}
\def\m@th{\mathsurround=0pt}
\mathchardef\bracell="0365
\def\upbrall{$\m@th\bracell$}
\def\underhat#1{\mathop{\vtop{\ialign{##\crcr
    $\hfil\displaystyle{#1}\hfil$\crcr
     \noalign
     {\kern1.5pt\nointerlineskip}
     \upbrall\crcr\noalign{\kern1pt
   }}}}\limits}
\usepackage{pgf}
\usepackage{tikz}
\usepackage{cite}
\usepackage{verbatim}
\usepackage[utf8]{inputenc}
\usetikzlibrary{arrows,automata}
\usetikzlibrary{positioning}
\def\theequation{\arabic{section}.\arabic{equation}}

\newcommand{\wh}{\widehat}
\newcommand{\wt}{\widetilde}

\def\hypotilde#1#2{\vrule depth #1 pt width 0pt{\smash{{\mathop{#2}
\limits_{\displaystyle\widetilde{}}}}}}
\def\hypohat#1#2{\vrule depth #1 pt width 0pt{\smash{{\mathop{#2}
\limits_{\displaystyle\widehat{}}}}}}

\newcommand{\bblu}{\begin{color}{blue}}
\newcommand{\bred}{\begin{color}{red}}
\newcommand{\ecl}{\end{color}}

\newcommand{\bLam}{\boldsymbol{\Lambda}}

\newcommand{\be}{\begin{equation}}
\newcommand{\ee}{\end{equation}}
\newcommand{\bea}{\begin{eqnarray}}
\newcommand{\eea}{\end{eqnarray}}
\newcommand{\bse}{\begin{subequations}}
\newcommand{\ese}{\end{subequations}}
\newcommand{\nn}{\nonumber}

\begin{document}

\def\theequation{\arabic{section}.\arabic{equation}}

\newtheorem{thm}{Theorem}[section]
\newtheorem{lem}{Lemma}[section]
\newtheorem{defn}{Definition}[section]
\newtheorem{ex}{Example}[section]
\newtheorem{rem}{}
\newtheorem{criteria}{Criteria}[section]
\newcommand{\ra}{\rangle}
\newcommand{\la}{\langle}
\makeatletter
\newcommand*{\rom}[1]{\expandafter\@slowromancap\romannumeral #1@}
\makeatother
\title{\textbf{ \textsf{On the Lagrangian 1-Form Structure of the Hyperbolic Calogero-Moser System}}}
\author{\\\\Umpon Jairuk$^\dagger$, Sikarin Yoo-Kong$^{\dagger,+,* } $ and Monsit Tanasittikosol$^{\dagger} $ \\
\small $^\dagger $\emph{Theoretical and Computational Physics (TCP) Group, Department of Physics,}\\ 
\small \emph{Faculty of Science, King Mongkut's University of Technology Thonburi, Thailand, 10140.}\\
\small $^+$\emph{Theoretical and Computational Science Center (TaCS),}\\ \small\emph{Faculty of Science, King Mongkut's University of Technology Thonburi, Thailand, 10140.}\\
\small $^* $ \emph{Ratchaburi Campus, King Mongkut's University of Technology Thonburi, Thailand, 70510.}\\
}
\maketitle
\abstract
In this work, we present another example of the Lagrangian 1-form structure for the hyperbolic Calogero-Moser system both in discrete-time level and continuous-time level. The discrete-time hyperbolic Calogero-Moser system is obtained by considering pole-reduction of the semi-discrete Kadomtsev-Petviashvili (KP) equation. The key relation called the discrete-time closure relation is directly obtained from the compatibility between the temporal Lax matrices. The continuous-time hierarchy of the hyperbolic Calogero-Moser system is obtained through two successive continuum limits. The continuous-time closure relation, which is a consequence of continuum limits on the discrete-time one, is also shown to hold.
\section{Introduction}\label{intro}
\setcounter{equation}{0}
The Lagrangian multiform structure has become one of the main research topics in the integrable systems after pioneer works were initiated by Lobb and Nijhoff \cite{SF1,SF2,SF3}. In these works, the discrete Lagrangian 2-form and 3-form for the systems with infinite degrees of freedom had been shown to possess a remarkable property called the \emph{closure relation} resulting from the variational principle on the space of independent variables. Soon later, the Lagrangian 1-form structure had been studied through the the systems with finite degrees of freedom namely the rational Calogero-Moser system \cite{Sikarin1} and the rational Ruijsenaar-Schneider system \cite{Sikarin2} by one of the authors. In these works, the discrete-time Lagrangians had been also shown to possess an intriguing property called the closure relation which guarantees the invariance of the action under local deformation of the curve on the space of independent variables. Then the continuum limits had been considered to obtain the hierarchy of the systems in the continuous-time case and continuous-time Lagrangians satisfied the closure relation. From this series of works on the Lagrangian multiform structure, we may conclude that there exists a Lagrangian analogue of the Liouville's integrability \cite{Babelon}. Later on, a number of works in this direction of research have been continuously published \cite{LAMS,LAMS1,LAMS2,LAMS3,LAMS4,LAMS5,Umpon}.
\\
\\
In the present paper, we report the result of study further more on the Lagrangian 1-form structure of the Calogero-Moser type systems, or more specifically on the hyperbolic Calogero-Moser system. In section \ref{DTFLOWS}, the two compatible discrete-time hyperbolic Calogero-Moser systems will be obtained from the semi-discrete Kadomtsev-Petviashvili (KP) equation through the pole-reduction method. The discrete Lagrangians are also established and the closure relation is directly obtained via the connection between the temporal Lax matrix and the Lagrangian. In section \ref{skewlimit}, the continuum limit will be performed on one of discrete-time variable resulting to a hierarchy of the semi-continuous hyperbolic Calogero-Moser system. In section \ref{fullLIMIT}, the remaining discrete-time variable is converted to the continuous-time variables leading to a hierarchy of the continous-time hyperbolic Calogero-Moser system. The summary will be provided in the last section together with some remarks.
\section{The discrete-time flows}\label{DTFLOWS}
\begin{figure}[h]
\begin{center}
\subfigure[Horizontal discrete-time steps. ]{
\begin{tikzpicture}[scale=0.45]
 \draw[->] (0,0) -- (13,0) node[anchor=west] {$n$};
 \draw[->] (0,0) -- (0,12) node[anchor=south] {$m$};
 \draw[thick,dashed,black] (2,7)--(7,7)--(12,7);
 \fill (2,7) circle (0.1);
 \fill (7,7) circle (0.1);
 \fill (12,7) circle (0.1);
 \draw (2,6.7) node[anchor=north] {$\hypotilde 0 {\boldsymbol x}$};
 \draw (7,7) node[anchor=north] {${{\boldsymbol x}}$};
 \draw (12,7) node[anchor=north] {${{\wt{\boldsymbol x}}}$};
 \end{tikzpicture}}
\subfigure[Vertical discrete-time steps.]{
\begin{tikzpicture}[scale=0.45]
\draw[->] (0,0) -- (13,0) node[anchor=west] {$n$};
\draw[->] (0,0) -- (0,12) node[anchor=south] {$m$};
\draw[thick,dashed,black] (7,2)--(7,7)--(7,12);
 \fill (7,2) circle (0.1);
 \fill (7,7) circle (0.1);
 \fill (7,12) circle (0.1);
 \draw (7,2) node[anchor=west] {$\hypohat 0 {\boldsymbol x}$};
 \draw (7,7) node[anchor=west] {${{\boldsymbol x}}$};
 \draw (7,12) node[anchor=west] {${{\wh{\boldsymbol x}}}$};
\end{tikzpicture}}
\end{center}
\caption{Discrete-time evolutions along the horizontal and vertical directions on the space of independent variables.}\label{HL}
\end{figure}
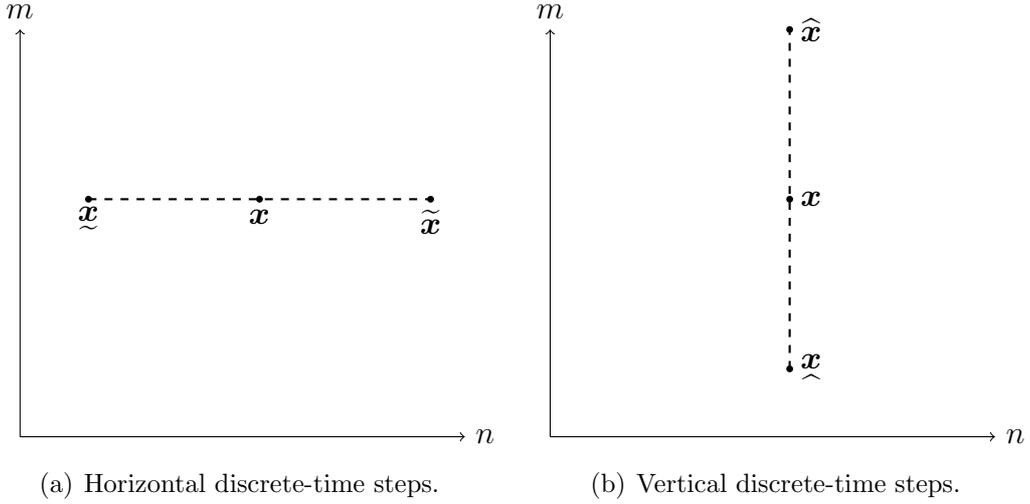

\begin{figure}[h]
\begin{center}
\subfigure[]{
\begin{tikzpicture}[scale=0.45]
 \draw[thick,dashed,black] (7,12)--(7,7)--(12,7);
 \fill (7,12) circle (0.1);
 \fill (7,7) circle (0.1);
 \fill (12,7) circle (0.1);
 \draw (7,12) node[anchor=west] {$\wh {\boldsymbol x}$};
 \draw (7,7) node[anchor=north] {${{\boldsymbol x}}$};
 \draw (12,7) node[anchor=north] {${{\wt{\boldsymbol x}}}$};
 \end{tikzpicture}}
\subfigure[]{
\begin{tikzpicture}[scale=0.45]
\draw[thick,dashed,black] (7,2)--(7,7)--(2,7);
 \fill (7,2) circle (0.1);
 \fill (7,7) circle (0.1);
 \fill (2,7) circle (0.1);
 \draw (7,2) node[anchor=west] {$\hypohat 0 {\boldsymbol x}$};
 \draw (7,7) node[anchor=west] {${{\boldsymbol x}}$};
 \draw (2,6.7) node[anchor=north] {$\hypotilde 0 {\boldsymbol x}$};
\end{tikzpicture}}
\subfigure[]{
\begin{tikzpicture}[scale=0.45]
\draw[thick,dashed,black] (7,12)--(7,7)--(2,7);
 \fill (7,12) circle (0.1);
 \fill (7,7) circle (0.1);
 \fill (2,7) circle (0.1);
 \draw (7,12) node[anchor=west] {$\wh {\boldsymbol x}$};
 \draw (7,7) node[anchor=west] {${{\boldsymbol x}}$};
 \draw (2,6.7) node[anchor=north] {$\hypotilde 0 {\boldsymbol x}$};
\end{tikzpicture}}
\subfigure[]{
\begin{tikzpicture}[scale=0.45]
\draw[thick,dashed,black] (7,2)--(7,7)--(12,7);
 \fill (7,2) circle (0.1);
 \fill (7,7) circle (0.1);
 \fill (12,7) circle (0.1);
 \draw (7,2) node[anchor=west] {$\hypohat 0 {\boldsymbol x}$};
 \draw (7,7) node[anchor=east] {${{\boldsymbol x}}$};
 \draw (12,7) node[anchor=north] {$\wt {\boldsymbol x}$};
\end{tikzpicture}}
\end{center}
\caption{Discrete-time evolutions around the corner.}\label{HL1}
\end{figure}
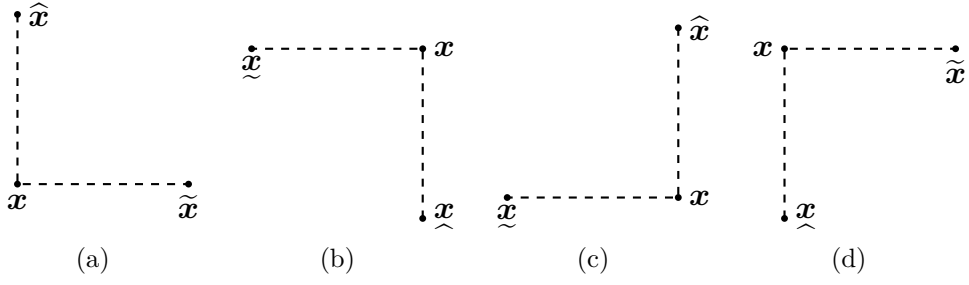

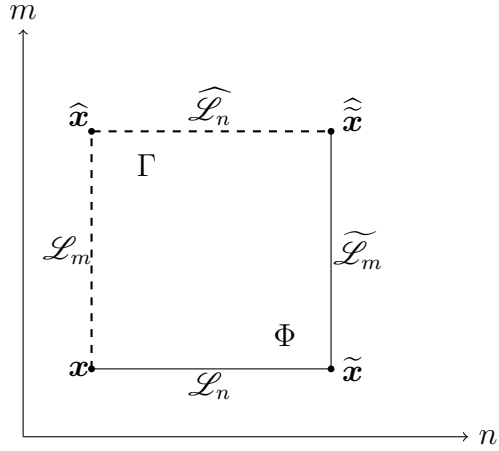
\begin{figure}[h]
\begin{center}
\begin{tikzpicture}[scale=0.45]
\draw[->] (0,0) -- (13,0) node[anchor=west] {$n$};
\draw[->] (0,0) -- (0,12) node[anchor=south] {$m$};
 \draw[thick,dashed,black] (2,2)--(2,9)--(9,9);
 \fill (2,2) circle (0.1);
 \fill (2,9) circle (0.1);
 \fill (9,2) circle (0.1);
 \fill (9,9) circle (0.1);
 \draw (1,2) node[anchor=west] {$\boldsymbol x$};
\draw (9,2) node[anchor=west] {$\wt{\boldsymbol x}$};
 \draw (1,9.5) node[anchor=west] {$\wh{\boldsymbol x}$};
 \draw (9,9.5) node[anchor=west] {$\wh{\wt{\boldsymbol x}}$};
 \draw (3,8)node[anchor=west] {$\Gamma$};
 \draw (7,3)node[anchor=west] {$\Phi$};
 \draw[black] (2,2) --(9,2)  -- (9,9);
\draw (4.5,9.75) node[anchor=west] {$\wh{\mathscr{L}}_{n}$};
\draw (0.25,5.5) node[anchor=west] {$\mathscr{L}_{m}$};
\draw (4.5,1.5) node[anchor=west] {$\mathscr{L}_{n}$};
\draw (8.75,5.5) node[anchor=west] {$\wt{\mathscr{L}}_{m}$};
\end{tikzpicture}
\end{center}
\caption{The local variation of the discrete curve on the space of independent variables.}\label{C3}
\end{figure}

\setcounter{equation}{0} 
In this section, we begin to construct the discrete-time Hyperbolic Calogero-Moser system by proceeding the same method provided in \cite{FrankCM,FrankCM2, Sikarin1}.
\\\\
\textbf{Pole-reduction}: We start to consider the semi-discrete Kadomtsev-Petviashvili (KP) equation given by
\begin{eqnarray} 
\partial_ \xi(\wh{u}-\wt{u})&=& (p-q+\wh{u}-\wt{u})(u+\wh{\wt {u}}-\wh{u}-\wt{u})\;, \label{SemiKP} 
\end{eqnarray}
where $p$ and $q$ are two lattice parameters. The variable $u$ is the classical field variable which depends on two discrete variables $(n,m)$ and a continuous variable $\xi$: $u\equiv u(n,m,\xi)$. The notations $\wt{u}=u(n+1,m,\xi)$ and $\wh{u}=u(n,m+1,\xi)$ are defined as the discrete-time evolution of the variable $u$ in $n-$ and $m-$directions, respectively. The combination of discrete-time evolutions is given by $\wh{\wt {u}}=u(n+1,m+1,\xi)$.
\\
\\
Eq. \eqref{SemiKP} is a consequence of the compatibility of the Lax pair given by
\begin{subequations}
\begin{eqnarray}
\wt{ \phi}&=&\phi_ \xi +(p+u-\wt{u})\phi \ , \label {SLP1}\\
\wh{ \phi}&=&\phi_ \xi +(q+u-\wh{u})\phi\;,\label {SLP2}
\end{eqnarray}
\end{subequations}
where, in the hyperbolic case, the variable $u$ can be chosen in the form 
\begin{eqnarray}
u(n,m,\xi) &= &\sum_{i=1}^N\coth{(\xi-x_i(n,m))}\; ,\label{u}
\end{eqnarray} 
and the plane wave function $\phi(n,m,\xi)$ is given in the form
\begin{eqnarray} \label{SLPP}
\phi &=& \left(1-\coth (\kappa) \sum_{i=1}^N b_i \coth (\xi -x_i) \right)(p+\sinh (\kappa))^n (q+\sinh (\kappa))^m e^{\sinh(\kappa)\xi}, \;
\end{eqnarray}
where $\kappa$ is a spectral parameter and the parameters $b_i \equiv b_i(n,m)  $ are yet to be determined. 
\\
\\
\textbf{{The n-flow}}: We substitute $u$ and $\phi$  into  the  Lax equation \eqref {SLP1} resulting to a coupled equation (see appendix A)
\begin{subequations}\label{LP} 
\begin{eqnarray}
(p+\sinh (\kappa))\boldsymbol b &=&\tanh (\kappa)\boldsymbol E +\boldsymbol L \boldsymbol b\ ,\label{LP1}\\
(p+\sinh (\kappa))\wt{\boldsymbol b} &=&\tanh (\kappa)\boldsymbol E +\boldsymbol M \boldsymbol b\;,\label{LP2}
\end{eqnarray}
\end{subequations}
where $\boldsymbol b=(b_1,b_2,...,b_N)^T$, $\boldsymbol E=(1,1,...,1)^T$. The Lax matrices $\boldsymbol L $ and $\boldsymbol M$ are given in the form 
\begin{subequations}
\begin{eqnarray}
\boldsymbol L &=& \sum_{i,j=1}^N \left(\coth(x_i-\wt{x}_j)-\coth(x_i-x_j)\right)E_{ii} - \sum_{j \ne i}^N \coth(x_i-x_j)E_{ij}\;, \label{MHCM} \\
\boldsymbol M &=& -\sum_{i,j=1}^N\coth(\wt x_i-x_j)E_{ij}\;. \label{MHCM}
\end{eqnarray}
\end{subequations}
The notation $E_{ij}$ is the element index of the matrix. In addition, the compatibility of equation \eqref{LP1} and  \eqref{LP2} leads to
\begin{eqnarray}
(\wt{\boldsymbol L}\boldsymbol M-\boldsymbol M\boldsymbol L) \boldsymbol b + \tanh(\kappa)(\wt{\boldsymbol L}-\boldsymbol M)\boldsymbol E=0\; ,\label{LPP}
\end{eqnarray}
which we obtain
\begin{subequations}\label{DLP} 
\begin{eqnarray}
\wt{\boldsymbol L}\boldsymbol M-\boldsymbol M\boldsymbol L&=&0\ ,\label{DLP1}\\
(\wt{\boldsymbol L}-\boldsymbol M)\boldsymbol E&=&0\ .\label{DLP2}
\end{eqnarray}
\end{subequations}
Both \eqref{DLP1} and \eqref{DLP2} produce
\begin{eqnarray}
\sum_{j=1}^N\left(\coth (x_i-\wt{x}_j)+ \coth (x_i-\hypotilde 0 {x}_j)\right)-2\sum_{j=1,j\neq i}^N\coth (x_i-x_j)=0\;,\label{EQHCM1}
\end{eqnarray}
which is the discrete-time equation of motion of the hyperbolic Calogero-Moser system in the $n-$direction, see figure \ref{HL}(a).
\\
\\
\textbf{{The m-flow}}: Next, if we substitute $u$ and $\phi$ into equation \eqref{SLP2} we obtain
\begin{subequations}\label{LPH} 
\begin{eqnarray}
(q+\sinh (\kappa))\boldsymbol b &=&\tanh (\kappa)\boldsymbol E +\boldsymbol K \boldsymbol b\ ,\label{LPH11}\\
(q+\sinh (\kappa))\wh{\boldsymbol b} &=&\tanh (\kappa)\boldsymbol E +\boldsymbol N \boldsymbol b\ , \label{LPH21}
\end{eqnarray}
\end{subequations}
where the Lax matrices $\boldsymbol K$ and $\boldsymbol N$ are in the form
\begin{subequations}\label{KHCMM}
\begin{eqnarray} 
\boldsymbol K&=&\sum_{i,j=1}^N\left(\coth{(x_i-\wh{x}_j)}- \coth{(x_i-x_j)}\right)E_{ii}-\sum_{j \ne i}^N\coth (x_i-x_j)E_{ij}\;,\\ \label{KHCM}
\boldsymbol N&=& -\sum_{i,j=1}^N\coth (\wh{x}_i-x_j)E_{ij}\;. \label{NHCM}
\end{eqnarray}
\end{subequations}
The compatibility between \eqref{LPH11} and  \eqref{LPH21} gives
\begin{eqnarray}
(\wh{\boldsymbol K}\boldsymbol N-\boldsymbol N\boldsymbol K) \boldsymbol b + \tanh(\kappa)(\wh{\boldsymbol K}-\boldsymbol N)\boldsymbol E=0\; ,\label{LPP}
\end{eqnarray}
resulting to
\begin{subequations}\label{DLPX} 
\begin{eqnarray}
\wh{\boldsymbol K}\boldsymbol N-\boldsymbol N\boldsymbol K&=&0\ ,\label{DLPX1}\\
(\wh{\boldsymbol K}-\boldsymbol N)\boldsymbol E&=&0\ .\label{DLPX2}
\end{eqnarray}
\end{subequations}
Both \eqref{DLPX1} and \eqref{DLPX2} produce
\begin{eqnarray}
\sum_{j=1}^N\left(\coth (x_i-\wh{x}_j)+ \coth (x_i-\hypohat 0 {x}_j)\right)-2\sum_{j=1,j\neq i}^N\coth (x_i-x_j)=0\;,\label{EQHCM2}
\end{eqnarray}
which is the discrete-time equation of motion of the hyperbolic Calogero-Moser system in the $m-$direction, see figure \ref{HL}(b).
\\
\\
\textbf{Constraints}:
The compatibilities between \eqref{LP1} and  \eqref{LPH11}
\begin{subequations}
\begin{eqnarray}
(p-q)\boldsymbol b&=&(\boldsymbol L-\boldsymbol K)\boldsymbol b\;,\label{CDLPH1}
\end{eqnarray}
and between \eqref{LP2} and \eqref{LPH21} 
 \begin{eqnarray}
(p-q)k\boldsymbol E &=&(\wh{\boldsymbol M}-\wt{\boldsymbol N})k\boldsymbol E+(\wh{\boldsymbol M}\boldsymbol N-\wt{\boldsymbol N}\boldsymbol M)\boldsymbol b\;, \label{CDLPH2}
\end{eqnarray}
\end{subequations}
which produce
\begin{subequations}\label{CDE} 
\begin{eqnarray}
p-q &= &\sum_{j=1}^N\left(\coth (x_i-\wt{x}_j) - \coth (x_i-\wh{x}_j)\right)\ ,\label{CDE1}\\
p-q &=&\sum_{j=1}^N\left(\coth (x_i-\hypohat 0 {x}_j) - \coth (x_i-\hypotilde 0 {x}_j)\right).\label{CDE2}
\end{eqnarray}
\end{subequations}
These two equations describe the discrete-time evolution of the system involving two different discrete directions shown in figures \ref{HL1}(a) and \ref{HL1}(b), respectively. Another two types of discrete-time curve shown in figures \ref{HL1}(c) and \ref{HL1}(d) can be obtained with the help of equations of motion \eqref{EQHCM1} and \eqref{EQHCM2}.
\\
\\
\textbf{{Exact solution}}: 
The exact solution for the hyperbolic Calogero-Moser system takes the form
\begin{eqnarray}\label{EXACT1}
e^{\boldsymbol{Y}(n,m)}&=& (p\boldsymbol{I}+\bLam)^{-n}(q\boldsymbol{I}+\bLam)^{-m}e^{\boldsymbol{Y}(0,0)}(p\boldsymbol{I}+\bLam)^{n}(q\boldsymbol{I}+\bLam)^{m}\nn\\
&&-\frac{n\bLam}{(p\boldsymbol{I}+\bLam)}-\frac{m\bLam}{(q\boldsymbol{I}+\bLam)}\;,
\end{eqnarray}
which its derivation is given in appendix B.
\\\\
\textbf{{The discrete action}}:
We find that the action for discrete curves given in the figure \ref{HL}(a) and \ref{HL}(b) are
\begin{subequations}
\begin{eqnarray}
{S}_{h}&=&\mathscr{L}_{(n)}(\boldsymbol{x},\wt{\boldsymbol{x}})+\mathscr{L}_{(n)}(\boldsymbol{x},{\hypotilde 0 {\boldsymbol{x}}})\;,\\ \label{dactionn}
{S}_{v}&=&\mathscr{L}_{(m)}(\boldsymbol{x},\wh{\boldsymbol{x}})+\mathscr{L}_{(m)}(\boldsymbol{x},{\hypohat 0 {\boldsymbol{x}}})\;,\label{dactionm}
\end{eqnarray}
\end{subequations}
where
\begin{subequations}
\begin{eqnarray}
\mathscr{L}_{(n)}(\boldsymbol{x},\wt{\boldsymbol{x}})&=&-\sum_{i,j=1}^N\ln\left|\sinh (x_i-\wt{x}_j)\right|+\frac{1}{2}\sum_{j\ne i}^N\left(\ln\left|\sinh (x_i-x_j)\right|\right.\nn\\
&&+\left.\ln\left|\sinh (\wt{x}_i-\wt{x}_j)\right|\right)
+p(\Xi-\wt \Xi) \;,\label{1Lagn}\\
\mathscr{L}_{(m)}(\boldsymbol{x},\wh{\boldsymbol{x}})&=&-\sum_{i,j=1}^N\ln\left|\sinh (x_i-\wh{x}_j)\right|+\frac{1}{2}\sum_{j\ne i}^N\left(\ln\left|\sinh (x_i-x_j)\right|\right.\nn\\
&&+\left.\ln\left|\sinh (\wh{x}_i-\wh{x}_j)\right|\right)
+q(\Xi-\wh \Xi)\;,\label{2Lagm}
\end{eqnarray}
\end{subequations}
where $\Xi=\sum_i^Nx_i$. Then we perform the local variation of the discrete curves on the space of dependent variables resulting to
\begin{subequations}
\begin{eqnarray}
\delta S_{h}=0\;\;\Rightarrow\;\;\;\; \frac{\partial\mathscr{L}_{(n)}}{\partial{\wt{x}_i}}+\wt{\frac{\partial\mathscr{L}_{(n)}}{\partial{x_i}}}&=&0\ , \label{Eactionn} \\
\delta S_{v}=0\;\;\Rightarrow\;\; \frac{\partial\mathscr{L}_{(m)}}{\partial{\wh{x}_i}}+\wh{\frac{\partial\mathscr{L}_{(m)}}{\partial{x_i}}}&=&0 \;.\label{Eactionm}
\end{eqnarray}
\end{subequations}
 \eqref{Eactionn} and \eqref{Eactionm} are discrete-time Euler-Lagrange equations yielding the equations of motion \eqref{EQHCM1} and \eqref{EQHCM2}, respectively. In addition, we have other four actions relating two different discrete trajectories around the corner, shown in the figure \ref{HL1}, given by
\begin{subequations}
\begin{eqnarray}
S_{(a)}&=&\mathscr{L}_{(n)}(\wt{\boldsymbol{x}},\boldsymbol{x})+\mathscr{L}_{(m)}(\boldsymbol{x},\wh{\boldsymbol{x}})\;,\\
S_{(b)}&=&\mathscr{L}_{(n)}({\hypotilde 0 {\boldsymbol{x}}},\boldsymbol{x})+\mathscr{L}_{(m)}(\boldsymbol{x},{\hypohat 0 {\boldsymbol{x}}})\;,\\
S_{(c)}&=&\mathscr{L}_{(n)}({\hypotilde 0 {\boldsymbol{x}}},\boldsymbol{x})+\mathscr{L}_{(m)}(\boldsymbol{x},\wh{\boldsymbol{x}})\;,\\
S_{(d)}&=&\mathscr{L}_{(n)}(\wt{\boldsymbol{x}},\boldsymbol{x})+\mathscr{L}_{(m)}(\boldsymbol{x},{\hypohat 0 {\boldsymbol{x}}})\;.
\end{eqnarray}
\end{subequations}
The local variation gives us
\begin{subequations}
\begin{eqnarray}
\delta S_{(a)}=0&\Rightarrow&\frac{\partial\mathscr{L}_{(n)}}{\partial{\wt{x}_i}}+{\frac{\partial\mathscr{L}_{(m)}}{\partial{\wh x_i}}}=0\;,\\
\delta S_{(b)}=0&\Rightarrow&\frac{\partial\mathscr{L}_{(n)}}{\partial{\hypotilde 0 {{x}}_i}}+\frac{\partial\mathscr{L}_{(m)}}{\partial{\hypohat 0 {{x}}_i}}=0\;,\\
\delta S_{(c)}=0&\Rightarrow&\frac{\partial\mathscr{L}_{(n)}}{\partial{\hypotilde 0 {{x}}_i}}+{\frac{\partial\mathscr{L}_{(m)}}{\partial{\wh x_i}}}=0\;,\\
\delta S_{(d)}=0&\Rightarrow&\frac{\partial\mathscr{L}_{(n)}}{\partial{\wt{x}_i}}+\frac{\partial\mathscr{L}_{(m)}}{\partial{\hypohat 0 {{x}}_i}}=0\;,
\end{eqnarray}
\end{subequations}
which yield the constraint equations.
\\
\\
\textbf{The closure relation}: We may employ the existence of the relation between temporal Lax matrix and the Lagrangian \cite{Sikarin1} to establish the closure relation. We consider the compatibility between the matrix $\boldsymbol M$ and the matrix $\boldsymbol N$ given by
\begin{eqnarray} \label{MNNM}
\wh{\boldsymbol M} \boldsymbol N &=& \wt{\boldsymbol N} \boldsymbol M \;.
\end{eqnarray}
This equation can be rewritten in the form
\begin{eqnarray} \label{LOGMN}
\log |\det \wh{\boldsymbol M} | + \log \left|\det \boldsymbol N \right|  &=& \log |\det \wt{\boldsymbol N} | + \log \left|\det \boldsymbol M \right| \;.
\end{eqnarray}
Using
\begin{eqnarray}
\mathscr{L}_{(n)}(\boldsymbol x,\wt{\boldsymbol x})&=&\log \left| \det \boldsymbol M \right|+p(\Xi-\wt \Xi)\;,\label{1Lagndet}\\
\mathscr{L}_{(m)}(\boldsymbol x,\wh{\boldsymbol x})&=&\log \left| \det \boldsymbol N \right|+q(\Xi-\wh \Xi) \;,\label{2Lagmdet}
\end{eqnarray}
then \eqref{LOGMN} becomes
\begin{equation}\label{clo}
 \widehat{\mathscr{L}_{(n)}}(\boldsymbol{x},\widetilde{\boldsymbol{x}}) - \mathscr{L}_{(n)}(\boldsymbol{x},\widetilde{\boldsymbol{x}}) -
 \widetilde{\mathscr{L}_{(m)}}(\boldsymbol{x},\widehat{\boldsymbol{x}}) + \mathscr{L}_{(m)}(\boldsymbol{x},\widehat{\boldsymbol{x}}) = 0\;,
\end{equation}
with the condition $\sum_{i=1}^N\left(\wt x_i+\wh x_i-x_i-\wh{\wt {x}}_i\right)=0$ holds on the solution. Equation \eqref{clo} is the well known relation for Lagrangian 1-form called the ``discrete-time closure relation". Actually \eqref{clo} is a direct result of the variation of discrete curve on the space of independent discrete variables shown in figure \ref{C3}. The action of discrete curve of $\Gamma$ is
\begin{eqnarray}\label{AP}
S_{\Gamma}&=& \mathscr{L}_{(m)}(\boldsymbol{x},\wh{\boldsymbol{x}}) +  \widehat{\mathscr{L}_{(n)}}(\boldsymbol{x},\widetilde{\boldsymbol{x}})\;,
\end{eqnarray}
and the action of  discrete curve of $\Phi$ is
\begin{eqnarray} \label{APP}
S_{\Phi} &=&  \mathscr{L}_{(n)}(\boldsymbol{x},\widetilde{\boldsymbol{x}}) +  \widetilde{\mathscr{L}_{(m)}}(\boldsymbol{x},\wh{\boldsymbol{x}})\;.
\end{eqnarray}
Under the condition on the local variation: $\delta S = S_{\Gamma} -S_{\Phi}=0$, resulting to
\begin{equation}\label{VP}
\mathscr{L}_{(m)}(\boldsymbol{x},\wh{\boldsymbol{x}}) +  \widehat{\mathscr{L}_{(n)}}(\boldsymbol{x},\widetilde{\boldsymbol{x}}) - \mathscr{L}_{(n)}(\boldsymbol{x},\widetilde{\boldsymbol{x}}) -  \widetilde{\mathscr{L}_{(m)}}(\boldsymbol{x},\wh{\boldsymbol{x}}) = 0 \;,
\end{equation}
which is indeed \eqref {clo}. The closure relation tells us that the action remains the same under  ``local" deformation of the discrete curve on the space of independent variables.
\\
\section{The semi-continuous time flows}\label{skewlimit}
\setcounter{equation}{0} 
In this section, we are interested to perform a continuum limit on one of discrete variables. However, a naive limit will produce only the first flow of the system in the hierarchy \cite{FrankCM}. In order to produce the whole continuous-time hierarchy (next section), we may need to cook up these two discrete variables: $\mathsf N = n+m$ to get a new pair of discrete variables $(\mathsf N,m)$. With this change of variables, we have
\begin{eqnarray}\label{changingvariables}
x(n,m)&\mapsto& \mathrm x(\mathsf N,m)=:{\mathrm x},\nn\\
\widetilde x = x(n+1,m) &\mapsto & \mathrm x(\mathsf N+1,m)=:\bar{\mathrm x}\;, \nn\\
\widehat x = x(n,m+1) &\mapsto & \mathrm x(\mathsf N+1,m+1)=:\widehat{\bar{ {\mathrm x}}}\;,\nn\\ 
{\widehat{\widetilde x}} = x(n+1,m+1) &\mapsto & \mathrm x(\mathsf N+2,m+1)=:\widehat{\bar{\bar {\mathrm x}}}\;.\nn
\end{eqnarray}
The plane wave function can now be rewritten in the form
\begin{eqnarray}\label{SKLP}
  \phi(\mathsf N,m)&=&\left(1-\coth (\kappa) \sum_{i=1}^N b_i \coth (\xi -x_i) \right)\left (p+\sinh (\kappa)\right)^{\mathsf N} \left( 1-\frac{\varepsilon}{p+\sinh(\kappa)}\right)^me^{\sinh(\kappa)\xi}\;,\nn\\
\end{eqnarray}
where $\varepsilon=p-q$. In \cite{Sikarin1,Sikarin2,Umpon}, the limit such that $n\rightarrow-\infty$, $m\rightarrow+\infty$ and $\varepsilon\rightarrow 0$ are considered resulting to
\begin{eqnarray}\label{SKLP2}
  \phi(\mathsf N,\tau)&=&\left(1-\coth (\kappa) \sum_{i=1}^N \mathrm b_i \coth (\xi -\mathrm x_i) \right)\left (p+\sinh (\kappa)\right)^{\mathsf N} e^{\sinh(\kappa)\xi -\frac{\tau}{p+\sinh (\kappa)}}\;,
\end{eqnarray}
where $  b_i (n,m)\rightarrow \mathrm b_i (\mathsf N,\tau)$ and $\tau=\varepsilon m$. In this present work, we do a little twist by rewriting \eqref{SKLP} in the form
\begin{eqnarray}\label{SKLP3}
  \phi(\mathsf N,m)&=&\left(1-\coth (\kappa) \sum_{i=1}^N b_i \coth (\xi -x_i) \right)\left (p+\sinh (\kappa)\right)^{\mathsf N} e^{\sinh(\kappa)\xi+m\ln { \left(1 -\frac{\varepsilon}{p+\sinh (\kappa)} \right)}}\;,\nn\\
\end{eqnarray}
and use the Taylor series with respect to the variable $\varepsilon$ \begin{eqnarray}\label{SemiSLP}
\phi(\mathsf N,\tau_l)&=&\left(1-\coth (\kappa) \sum_{i=1}^N \mathrm b_i \coth (\xi -\mathrm x_i) \right)\left (p+\sinh (\kappa)\right)^{\mathsf N} e^{\sinh(\kappa)\xi+ \sum_{l=1} \ \mathrm \mu_l \tau_l} \;,
\end{eqnarray}
where $\tau_l = m{\varepsilon^l}$ and $  b_i (n,m)\rightarrow \mathrm b_i (\mathsf N,\tau_l)$. The variable $\mu_l$ is defined
\begin{eqnarray}
\mathrm \mu_l &=& -\frac {1}{l} \frac{1}{(p+\sinh (\kappa))^l} \;.
\end{eqnarray}
The reason that we generate many variables of $\tau$ because only  $(\mathsf N,\tau)$ are only enough to produce the first two flows in the hierarchy of the continuous-time hyperbolic Calogero-Moser system \cite{Sikarin1,Sikarin2,Umpon}. In order to get the third flow, we need $(\mathsf N,\tau_1,\tau_2)$ and to get the fourth flow, we need $(\mathsf N,\tau_1,\tau_2,\tau_3)$ and so on.
\\
\\
\textbf{{Lax pair}}:
Under changing of variables,  \eqref{SLP1} and \eqref{SLP2} become
\begin{subequations}
\begin{eqnarray}
\wt{ \phi}&=&\phi_ \xi +(p+u-\wt{u})\phi \ , \label {SSLP1}\\
\wh{\wt{ \phi}}&=&\phi_ \xi +(p-\varepsilon+u-\wh{\wt{u}})\phi\;,\label {SSLP2}
\end{eqnarray}
\end{subequations}
where $\wt{ \phi} = \phi(\mathsf N+1,m,\xi)$ and $\wh{\wt{ \phi}}=\wh{\phi}(\mathsf N+1,m+1,\xi)$.
We expand \eqref {SSLP2} with respect to the variable $\varepsilon$ to get
\begin{eqnarray}
\bar {\phi} +\varepsilon \frac{\partial \bar {\phi}}{\partial \tau_1} + \varepsilon^2\left( \frac{\partial\bar \phi}{\partial \tau_2}+\frac{1}{2}\frac{\partial^2 \bar\phi}{\partial \tau_1^2}\right)+...  &=& \phi_ \xi +\left(p- \varepsilon +u-\bar{u} - \varepsilon \frac{\partial \bar{u}}{\partial \tau_1} \right.\nn\\
&&\left.- \varepsilon^2\left( \frac{\partial \bar u}{\partial \tau_2}+\frac{1}{2}\frac{\partial^2 \bar u}{\partial \tau_1^2}\right)...\right)\phi\;.\label {SSLLPP}
\end{eqnarray}
We collect the equations in the following
\begin{subequations}\label{DEQQQ}
\begin{eqnarray}
&&\mathcal O(\varepsilon^0 ):\;\;\;\;\;\bar \phi= \phi_{\xi}+(p+u-\bar u)\phi\;,\label{phi0}\\
&&\mathcal O(\varepsilon^1 ):\;\;\;\;\;
\frac{\partial \bar \phi}{\partial \tau_1}=-\left( 1+\frac{\partial\bar u}{\partial \tau_1}\right)\phi
\;,\label{phi1}\\
&&\mathcal O(\varepsilon^2 ):\;\;\;\;\;
\frac{\partial \bar\phi}{\partial \tau_2}+\frac{1}{2}\frac{\partial^2 \bar\phi}{\partial \tau_1^2}=-\left( \frac{\partial \bar u}{\partial \tau_2}+\frac{1}{2}\frac{\partial^2 \bar u}{\partial \tau_1^2} \right)\phi
\;.\label{phi2}
\end{eqnarray}
\end{subequations}
\eqref{DEQQQ} contains only the first three relations from the expansion in \eqref{SSLLPP}. The higher terms can be obtained by just pushing forwards in the expansion.
\\\\
\emph{The $\mathsf N-$flow}: Using \eqref{SemiSLP} and the definition of the $u$, \eqref{phi0} gives us
\begin{subequations}\label{LP} 
\begin{eqnarray}
(p+\sinh (\kappa))\boldsymbol {\mathrm b} &=&\tanh (\kappa)\boldsymbol E +\boldsymbol {\mathrm L}\boldsymbol {\mathrm b}\ ,\label{LP12}\\
(p+\sinh (\kappa))\bar{\boldsymbol {\mathrm b}} &=&\tanh (\kappa)\boldsymbol E +\boldsymbol {\mathrm M} \boldsymbol {\mathrm b}\;,\label{LP22}
\end{eqnarray}
\end{subequations}
where the Lax matrices $\boldsymbol {\mathrm L} $ and $\boldsymbol {\mathrm M}$ are
\begin{subequations}
\begin{eqnarray}
\boldsymbol {\mathrm L} &=& \sum_{i,j=1}^N \left(\coth(x_i-\bar{x}_j)-\coth(x_i-x_j)\right)E_{ii} - \sum_{j \ne i}^N \coth(x_i-x_j)E_{ij}\;, \label{MHCM2} \\
\boldsymbol {\mathrm M} &=& -\sum_{i,j=1}^N\coth(\bar{ x}_i-x_j)E_{ij}\;. \label{MHCM2}
\end{eqnarray}
\end{subequations}
The compatibility between \eqref{LP12} and \eqref{LP22} produces again the discrete-time hyperbolic Calogero-Moser system
\begin{eqnarray}
\sum_{j=1}^N\left(\coth ( x_i-{\bar{  x}}_j)+\coth ( x_i-\underline x_j )\right)-2\sum_{j \ne i}^N\coth ( x_i- x_j) &=& 0 \;,\label{SKEWEQ1}
\end{eqnarray}
but with a new discrete-time variable $\mathsf N$.
\\
\\
\emph{The $\tau_1-$flow}: \eqref{phi1} gives us 
\begin{subequations}
\begin{eqnarray}
(p+\sinh(\kappa))\frac{\partial\bar{\boldsymbol {\mathrm b}}}{\partial\tau_1}&=&\bar{\boldsymbol {\mathrm b}}+{\boldsymbol {\mathrm A}}{\boldsymbol {\mathrm b}}\;,\label{b1}\\
(p+\sinh (\kappa))\bar{\boldsymbol {\mathrm b}} &=&\tanh (\kappa)\boldsymbol E +\boldsymbol {\mathrm M} {\boldsymbol {\mathrm b}}\;, \label{b2}\\
-1&=&\sum_{j=1}^N\frac{\partial \bar{\mathrm x}_j}{\partial\tau_1}\csch^2(\mathrm x_i-\bar{\mathrm x}_j)\;,\label{c1}
\end{eqnarray}
\end{subequations}
where the matrix $\boldsymbol {\mathrm A}$ is given in the form
\begin{equation}
\boldsymbol {\mathrm A}=\sum_{i,j=1}^N\frac{\partial \bar{\mathrm x}_i}{\partial\tau_1}\csch^2(\bar{\mathrm x}_i-\mathrm x_j)E_{ij}\;.
\end{equation}
Furthermore, we consider the compatibility between \eqref{b1} and \eqref{b2} resulting to \cite{Sikarin1}
\begin{subequations}
\begin{eqnarray}
-1&=&\sum_{j=1}^N\frac{\partial \underline{\mathrm x}_j}{\partial\tau_1}\csch^2(\mathrm x_i-\underline{\mathrm x}_j)\;,\label{c3}\\
0&=&\sum_{j=1}^N\left(\frac{\partial {\mathrm x}_j}{\partial\tau_1}\csch^2(\bar{\mathrm x}_i-\mathrm x_j)-
\frac{\partial {\mathrm x}_j}{\partial\tau_1}\csch^2(\underline{\mathrm x}_i-\mathrm x_j) \right)\;.\label{e1}
\end{eqnarray}
\end{subequations}
\eqref{c1} and \eqref{c3} are the constraints of the system which can be obtained by direct semi-continuum limit of the discrete-time equations of motion, later see \eqref{SKCE1} and \eqref{SKCEE2}. Using \eqref{c1} and \eqref{c3}, we can rewrite \eqref{e1} in the form
\begin{equation}
0=\sum_{j=1}^N\left(\frac{\partial \bar{\mathrm x}_j}{\partial\tau_1}\csch^2({\mathrm x}_i-\bar{\mathrm x}_j)-
\frac{\partial\underline {\mathrm x}_j}{\partial\tau_1}\csch^2({\mathrm x}_i-\underline{\mathrm x}_j) \right)\;,\label{e2}
\end{equation}
which is the equations of the motion associated with the $\tau_1-$flow, later see \eqref{DEQ1}.
\\
\\
\emph{The $\tau_2-$flow}: \eqref{phi2} gives us
\begin{subequations}
\begin{eqnarray}
(p+\sinh(\kappa))\left(\frac{\partial\bar{\boldsymbol {\mathrm b}}}{\partial\tau_2}+\frac{1}{2}\frac{\partial^2\bar{\boldsymbol {\mathrm b}}}{\partial\tau_1^2}\right)&=&\frac{\partial\bar{\boldsymbol {\mathrm b}}}{\partial\tau_1}+{\boldsymbol {\mathrm B}}{\boldsymbol {\mathrm b}}\;,\label{b12}\\
(p+\sinh (\kappa))\bar{\boldsymbol {\mathrm b}} &=&\tanh (\kappa)\boldsymbol E +\boldsymbol {\mathrm M} {\boldsymbol {\mathrm b}}\;,\label{b22}
\end{eqnarray}
\begin{eqnarray}
0&=&\sum_{j=1}^N\left(\left( \frac{\partial {\bar {\mathrm x}}_j}{\partial \tau_2}+ \frac{1}{2} \frac{\partial^2 {\bar {\mathrm x}}_j}{\partial \tau_1^2} \right)\csch^{2}(\mathrm x_i-{\bar{\mathrm x}}_j)+ \left(\frac{\partial {\bar {\mathrm x}}_j}{\partial \tau_1}\right)^2\frac{\coth(\mathrm x_i-{\bar{\mathrm x}}_j)}{\sinh^{2}(\mathrm x_i-{\bar{\mathrm  x}}_j)}  \right) \;,\label{c12}
\end{eqnarray}
\end{subequations}
where the matrix $\boldsymbol{\mathrm B}$ is given by
\begin{equation}
\boldsymbol{\mathrm B}=\sum_{i,j=1}^N\left(\left( \frac{\partial {\bar {\mathrm x}}_i}{\partial \tau_2}+ \frac{1}{2} \frac{\partial^2 {\bar {\mathrm x}}_i}{\partial \tau_1^2} \right)\csch^{2}(\bar{\mathrm x}_i-{{\mathrm x}}_j) -\left(\frac{\partial {\bar {\mathrm x}}_i}{\partial \tau_1}\right)^2\frac{\coth(\bar{\mathrm x}_i-{{\mathrm x}}_j)}{\sinh^{2}(\bar{\mathrm x}_i-{{\mathrm  x}}_j)}\right)E_{ij}\;.
\end{equation}
The compatibility between \eqref{b12} and \eqref{b22} gives 
\begin{eqnarray}
&&\sum_{j=1}^N\left(\left(-\frac{\partial {\underline {\mathrm x}_j}}{\partial \tau_2}+ \frac{1}{2} \frac{\partial^2 {\underline  {\mathrm x}_j}}{{\partial \tau_1^2}}\right)\csch^{2}(\mathrm x_i-{\underline  {\mathrm x}_j})+\left(\frac{\partial {\underline  {\mathrm x}}_j}{\partial \tau_1}\right)^{2}\frac{\coth(\mathrm x_i-{\underline {\mathrm x}_j})}{\sinh^{2}(\mathrm x_i-{\underline  {\mathrm x}}_j)}  \right) =0\;,\label{c23}\\
&&\sum_{j=1}^N\left( \left( \frac{\partial {\bar {\mathrm x}_j}}{\partial \tau_2} + \frac{1}{2} \frac{\partial^2 {\bar {\mathrm x}_j}}{\partial \tau_1^2} \right)\csch^{2}(\mathrm x_i-\bar{\mathrm  x}_j) + \left(\frac{\partial {\bar {\mathrm x}}_j}{\partial \tau_1}\right)^2 \frac{\coth(\mathrm x_i-{\bar{\mathrm x}}_j)}{ \sinh^{2}(\mathrm x_i-\bar{\mathrm  x}_j)} \right.\nn\\
&&\left.+\left(-\frac{\partial {\underline {\mathrm x}_j}}{\partial \tau_2}+ \frac{1}{2} \frac{\partial^2 {\underline  {\mathrm x}_j}}{{\partial \tau_1^2}}\right)\csch^{2}(\mathrm x_i-{\underline  {\mathrm x}_j})+\left(\frac{\partial {\underline  {\mathrm x}}_j}{\partial \tau_1}\right)^{2}\frac{\coth(\mathrm x_i-{\underline {\mathrm x}_j})}{\sinh^{2}(\mathrm x_i-{\underline  {\mathrm x}}_j)}   \right)=0\;.\label{DEQ21}
\end{eqnarray}
\eqref{c12} and \eqref{c23} are the constraint and \eqref{DEQ21} is the equations of motion of the system associated with the $\tau_2-$flow.
\\
\\
\textbf{{ Exact solution}}: The solution \eqref{EXACT1} can be rewritten in the form
\begin{subequations}\label{EXACTSKEW}
\begin{eqnarray}\label{EXACTSKEW1}
\mathsf e^{{\boldsymbol Y}(n,m)}\mapsto \mathsf e^{{\boldsymbol Y}(\mathsf {N},m)}&=& e^{-\mathsf N\ln (p+\boldsymbol{\Lambda})}e^{-m\ln \left(1+\frac{\varepsilon}{p+\boldsymbol{\Lambda}}\right)}e^{\boldsymbol  Y(0,0)}e^{m\ln \left(1+\frac{\varepsilon}{p+\boldsymbol{\Lambda}}\right)}e^{\mathsf N\ln (p+\boldsymbol{\Lambda})}\nn\\
&&-\frac{\mathsf N}{p+\boldsymbol{\Lambda}} -\frac{m\varepsilon}{(p+\boldsymbol{\Lambda})^2}-\frac{m\varepsilon^2}{(p+\boldsymbol{\Lambda})^3}-...\;.
\end{eqnarray}
With the definition of $\tau_l$, we have
\begin{eqnarray} \label{EXACTSKEW2}
\mathsf e^{{\boldsymbol Y}(\mathsf N,m)}\mapsto \mathsf e^{{\boldsymbol Y}( \mathsf N,\tau_l)}&=& e^{-\mathsf N\ln \left(1+\frac{\boldsymbol{\Lambda}}{p}\right)}e^{\sum_l \tau_l \left(\frac{1}{l(p+\boldsymbol{\Lambda})^l}\right)}e^{\boldsymbol  Y(0,0)}e^{-\sum_l \tau_l \left(\frac{1}{l(p+\boldsymbol{\Lambda})^l}\right)}e^{\mathsf N\ln \left(1+\frac{\boldsymbol{\Lambda}}{p}\right)}\nn \\
&&-\frac{\mathsf N}{p+\boldsymbol{\Lambda}} -\frac{\tau_1}{(p+\boldsymbol{\Lambda})^2}-\frac{\tau_2}{(p+\boldsymbol{\Lambda})^3}-...\;,
\end{eqnarray}
\end{subequations}
which is the exact solution in the semi-continuous time case.
\\
\\
\textbf{{Equations of motion}}: Under changing of variables, the equations of motion \eqref{EQHCM2} becomes
\begin{eqnarray}
\sum_{j=1}^N\left(\coth (\mathrm x_i-\wh{\bar{\mathrm{  x}}}_j)+\coth (\mathrm x_i-\underline {\hypohat 0 { \mathrm{x}}}_j )\right)-2\sum_{j \ne i}^N\coth ( \mathrm x_i- \mathrm x_j) &=& 0 \;.\label{SKEWEQ}
\end{eqnarray}
Using the relations
\begin{subequations}\label{OPEE}
\begin{eqnarray}\label{OPER}
\wh{ \mathrm{x}}&=&\left(e^{\sum_{l} \varepsilon^l\frac{\partial }{\partial \tau_l} }\right)\mathrm{x}\nn\\
&=&\mathrm{x}+\varepsilon \frac{\partial \mathrm{x} }{\partial \tau_1}+\varepsilon^2 \left(\frac{1}{2} \frac{\partial^2 \mathrm{x}}{\partial \tau_1^2} +  \frac{\partial \mathrm{x}}{\partial \tau_2}\right)+\varepsilon^3 \left(\frac{1}{6} \frac{\partial^3 \mathrm{x}}{\partial \tau_1^3} +  \frac{\partial^2 \mathrm{x}}{\partial \tau_1\tau_2} + \frac{\partial \mathrm{x}}{\partial \tau_3}\right) +...,\;
\end{eqnarray}
\begin{eqnarray} \label{OPERR}
\hypohat 0 { \mathrm{x}} &=&\left(e^{-\sum_{l} \varepsilon^l\frac{\partial }{\partial \tau_l} }\right)\mathrm{x}\nn\\
&=&\mathrm{x}-\varepsilon \frac{\partial \mathrm{x} }{\partial \tau_1}-\varepsilon^2 \left(-\frac{1}{2} \frac{\partial^2 \mathrm{x}}{\partial \tau_1^2} +  \frac{\partial \mathrm{x}}{\partial \tau_2}\right)-\varepsilon^3 \left(\frac{1}{6} \frac{\partial^3 \mathrm{x}}{\partial \tau_1^3} - \frac{\partial^2 \mathrm{x}}{\partial \tau_1\tau_2} + \frac{\partial \mathrm{x}}{\partial \tau_3}\right) +....\;
\end{eqnarray}
\end{subequations}
Employing the above relations and expanding \eqref{SKEWEQ} with respect to the variable $\varepsilon$, we find that
\begin{subequations}\label{DEQ}
\begin{eqnarray}
&&\mathcal O(\varepsilon^0 ):\;\;\;\;\;\sum_{j=1}^N\left(\coth (\mathrm  x_i-\bar {\mathrm  {x}}_j)+\coth (\mathrm x_i-\underline  {\mathrm {x}}_j )\right)-2\sum_{j \ne i}^N\coth (\mathrm x_i-\mathrm x_j) = 0\;,\label{DEQ0}\\
&&\mathcal O(\varepsilon^1 ):\;\;\;\;\;\sum_{j=1}^N\left(\frac{\partial \bar{\mathrm {x}}_j}{\partial \tau_1} \csch^2(\mathrm {x}_i-\bar{\mathrm {x}}_j)-\frac{\partial \underline{\mathrm {x}}_j}{\partial \tau_1} \csch^2(\mathrm {x}_i-\underline {\mathrm {x}}_j)\right)=0\;,\label{DEQ1}\\
&&\mathcal O(\varepsilon^2 ):\;\;\;\;\;\sum_{j=1}^N\left( \left( \frac{\partial {\bar {\mathrm x}_j}}{\partial \tau_2} + \frac{1}{2} \frac{\partial^2 {\bar {\mathrm x}_j}}{\partial \tau_1^2} \right)\csch^{2}(\mathrm x_i-\bar{\mathrm  x}_j) + \left(\frac{\partial {\bar {\mathrm x}}_j}{\partial \tau_1}\right)^2 \frac{\coth(\mathrm x_i-{\bar{\mathrm x}}_j)}{ \sinh^{2}(\mathrm x_i-\bar{\mathrm  x}_j)} \right.\nn\\
&&\left.\;\;\;\;\;\;\;\;\;\;\;\;\;\;\;+\left(-\frac{\partial {\underline {\mathrm x}_j}}{\partial \tau_2}+ \frac{1}{2} \frac{\partial^2 {\underline  {\mathrm x}_j}}{{\partial \tau_1^2}}\right)\csch^{2}(\mathrm x_i-{\underline  {\mathrm x}_j})+\left(\frac{\partial {\underline  {\mathrm x}}_j}{\partial \tau_1}\right)^{2}\frac{\coth(\mathrm x_i-{\underline {\mathrm x}_j})}{\sinh^{2}(\mathrm x_i-{\underline  {\mathrm x}}_j)}   \right)=0\;. \nn\\\label{DEQ2} 
\end{eqnarray}
\end{subequations}
Here we provide only the first three equations in the expansion which are identical with what we had derived before from the Lax equations. The equation \eqref{DEQ0} is the equation of motion of the hyperbolic Calogero-Moser system with the new discrete variable $\mathsf N$. The equations \eqref{DEQ1} and \eqref{DEQ2} are also the equation of motion associating with the variable $\tau_1$ and $\tau_2$, respectively. Then the system of equations \eqref{DEQ} forms a hierarchy of the semi-continuous time hyperbolic Calogero-Moser system.
\\
\\
\textbf{Constriants}: We proceed the same steps of the semi-continuous limit on \eqref{CDE1} leading to
\begin{subequations}
\begin{eqnarray}\label{SKCE}
&&\mathcal O(\varepsilon^1 ):\;\;\;\;\;\sum_{j=1}^N \frac{\partial {\bar {\mathrm x}_j}}{\partial \tau_1}\csch^{2}(\mathrm x_i-{\bar{\mathrm  x}}_j)=-1\;,\label{SKCE1}\\
&&\mathcal O(\varepsilon^2 ):\;\;\;\;\;\nn\\
&&\sum_{j=1}^N\left(\left( \frac{\partial {\bar {\mathrm x}}_j}{\partial \tau_2}+ \frac{1}{2} \frac{\partial^2 {\bar {\mathrm x}}_j}{\partial \tau_1^2} \right)\csch^{2}(\mathrm x_i-{\bar{\mathrm x}}_j)+ \left(\frac{\partial {\bar {\mathrm x}}_j}{\partial \tau_1}\right)^2\frac{\coth(\mathrm x_i-{\bar{\mathrm x}}_j)}{\sinh^{2}(\mathrm x_i-{\bar{\mathrm  x}}_j)}  \right) =0\;.\label{SKCE11}
\end{eqnarray}
\end{subequations}
The constraint equation \eqref{CDE2} produces
\begin{subequations}
\begin{eqnarray}\label{SKCE}
&&\mathcal O(\varepsilon^1 ):\;\;\;\;\;\sum_{j=1}^N \frac{\partial \underline {\mathrm x}_j}{\partial \tau_1}\csch^{2}{(\mathrm x_i-\underline {\mathrm x}_j)}=-1\;,\label{SKCEE2}\\
&&\mathcal O(\varepsilon^2 ):\;\;\;\;\;\nn\\
&&\sum_{j=1}^N\left(\left(-\frac{\partial {\underline {\mathrm x}_j}}{\partial \tau_2}+ \frac{1}{2} \frac{\partial^2 {\underline  {\mathrm x}_j}}{{\partial \tau_1^2}}\right)\csch^{2}(\mathrm x_i-{\underline  {\mathrm x}_j})+\left(\frac{\partial {\underline  {\mathrm x}}_j}{\partial \tau_1}\right)^{2}\frac{\coth(\mathrm x_i-{\underline {\mathrm x}_j})}{\sinh^{2}(\mathrm x_i-{\underline  {\mathrm x}}_j)}  \right) =0\;.\label{SKCEE22}
\end{eqnarray}
\end{subequations}
Using \eqref{SKCE11} and \eqref{SKCEE2}, the equation of motion \eqref{DEQ2} can be further simplified to
\begin{eqnarray}\label{DEQQ}
\sum_{j=1}^N\left( \frac{\partial {\bar {\mathrm x}_j}}{\partial \tau_2}\csch^{2}({\mathrm x}_i-\bar {\mathrm  x}_j) - \frac{\partial {\bar  {\mathrm x}_i}}{\partial \tau_1}\frac{\partial {\mathrm x_j}}{\partial \tau_1}\frac{\coth({{\mathrm x}}_i-\bar {\mathrm x}_j)}{\sinh^{2}({{\mathrm x}}_i-\bar{ \mathrm x}_j)} \right)=0\;.
\end{eqnarray}
\\
\\
\textbf{{Lagrangians}}: The Lagrangian \eqref{2Lagm} in terms of the new variables $(\mathsf {N},m)$:
\begin{eqnarray}
\mathscr{L}_{(m)}&=&\sum_{i,j=1}^N\ln \left|\sinh( \mathrm  x_i-\wh{\bar{\mathrm  x}}_j)\right| + \frac{1}{2} \sum_{j\neq i}^N \left(\ln\left|\sinh(\mathrm  x_i- \mathrm  x_j)\right|\right.\nn\\
&&\left.+\ln\left|\sinh(\wh{\bar{\mathrm  x}}_i -\wh{\bar{\mathrm  x}}_j) \right|\right)
+p\Xi - p\wh{\bar{\Xi}} - \varepsilon\Xi - \varepsilon\wh{\bar{\Xi}} \;.\label{Lagm2}
\end{eqnarray}
We then expand with respect to the variable $\varepsilon$ resulting to
\begin{eqnarray}
\mathscr{L}_{(m)} \mapsto  \varepsilon^0 \mathcal{L}_{(\mathsf N)}+ \varepsilon^1 \mathcal{L}_{(\tau_1)}+  \varepsilon^2\mathcal{L}_{(\tau_2)}+...\;,\label{2Lagmv}
\end{eqnarray}
where 
\begin{subequations}\label{LTA}
\begin{eqnarray}\label{LTAUN}
\mathcal{L}_{(\mathsf N)} &=&-\sum_{i,j=1}^N\ln\left|\sinh(\mathrm x_i-{\bar{\mathrm  x}}_j)\right| + \frac{1}{2} \sum_{j\neq i}^N \ln\left|\sinh(\mathrm x_i-\mathrm x_j)\sinh ({\bar{\mathrm  x}}_i-{\bar{\mathrm  x}}_j)\right| + p(\Xi -\bar{\Xi}) \;,\label{LTAU0}\nn\\  \\
\mathcal{L}_{(\tau_1)}&=&\sum_{i,j=1}^N\frac{\partial {\bar{\mathrm  x}}_j}{\partial \tau_1}\coth(\mathrm x_i-{\bar{\mathrm  x}}_j) + \sum_{j\neq i}^N\frac{\partial {\bar{\mathrm  x}}_j}{\partial \tau_1}\coth({\bar{\mathrm  x}}_i-{\bar{\mathrm  x}}_j) - p\frac{\partial\bar{ \Xi}}{\partial \tau_1}+\bar{\Xi } - \Xi  \; , \label{LTAU1}\\
\mathcal{L}_{(\tau_2)}&=& \sum_{i,j=1}^N\left(\left(\frac{\partial {\bar{\mathrm  x}}_j}{\partial \tau_2}+\frac{1}{2}\frac{\partial^2 {\bar{\mathrm  x}}_j}{\partial \tau_1^2}\right)
\coth(\mathrm x_i-{\bar{\mathrm  x}}_j) +\frac{1}{2}\left(\frac{\partial {\bar{\mathrm  x}}_j}{\partial \tau_1}\right)^2 \csch^2(\mathrm x_i-{\bar{\mathrm  x}}_j)\right)\nn\\
&&-\sum_{j\neq i}^N\left(\left(\frac{\partial {\bar{\mathrm  x}}_j}{\partial \tau_2} + \frac{1}{2}\frac{\partial^2 {\bar{\mathrm  x}}_j}{\partial \tau_1^2}\right) \coth({\bar{\mathrm  x}}_i-{\bar{\mathrm  x}}_j) + \frac{1}{2}\left(\frac{\partial {\bar{\mathrm  x}}_j}{\partial \tau_1}\right)^2 \csch^2({\bar{\mathrm  x}}_i-{\bar{\mathrm  x}}_j)\right)\nn\\
&&+\sum_{j\neq i}^N\frac{1}{2}\frac{\partial {\bar{\mathrm  x}}_i}{\partial \tau_1}\frac{\partial {\bar{\mathrm  x}}_j}{\partial \tau_1}\csch^2({\bar{\mathrm  x}}_i - {\bar{\mathrm  x}}_j)
-p\left(\frac{1}{2}\frac{\partial^2 \bar{\Xi}}{\partial \tau_1^2}+ \frac{\partial \bar{\Xi}}{\partial \tau_2}\right) + \frac{\partial \bar{\Xi}}{\partial \tau_1} \; . \label{LTAU2} 
\end{eqnarray}
\end{subequations}
These Lagrangians produce the equations of motion \eqref{DEQ1}, \eqref{DEQ2} and \eqref{DEQQ}, respectively. This can be seen by substituting the Lagrangians in the following Euler-Lagrange equations
\begin{subequations}\label{LL1}
\begin{eqnarray}
\frac{\partial\mathcal{L}_{(\mathsf N)}}{\partial \mathrm  x}+\underline{\frac{\partial\mathcal{L}_{(\mathsf N)}}{\partial \bar{\mathrm  x}}}&=&0\;,\\
\frac{\partial \mathcal{L}_{(\tau_1)}}{\partial \mathrm  {x}} - \frac{d}{d\tau_1}\left(\frac{\partial \mathcal{L}_{(\tau_1)}}{\partial \left(\frac{\partial\mathrm  {x}}{\partial \tau_1}\right)}\right)&=&0\;,\\
\frac{\partial \mathcal{L}_{(\tau_2)}}{\partial \mathrm  {x}} - \frac{d}{d\tau_2}\left(\frac{\partial \mathcal{L}_{(\tau_2)}}{\partial \left(\frac{\partial\mathrm  {x}}{\partial \tau_2}\right)}\right)&=&0\;.
\end{eqnarray}
\end{subequations}
From \eqref{LL1}, we see that $\{\mathcal{L}_{(\mathsf N)},\mathcal{L}_{(\tau_1)},\mathcal{L}_{(\tau_2)},... \}$ form the Lagrangian hierarchy in the semi-continuous time case.
\section{The continuous-time flows}\label{fullLIMIT}
\begin{figure}[h]
\begin{center}
\begin{tikzpicture}[scale=0.6]
 \draw[->] (0,0) -- (9,0) node[anchor=west] {$t_{2}$};
 \draw[->] (0,0) -- (0,7) node[anchor=south] {$t_{3}$};
 \fill (1,1) circle (0.2) node[anchor=west]{$(t_2(s_0),t_3(s_0))$};
 \draw (0.02,4) node[anchor=north west] {$\Gamma_2$};
 \draw (2,5.9) node[anchor=north west] {$\Gamma_1$};
 \draw[thick](1,1)--(1,5)--(5,5);
 \draw (3,4) node[anchor=north west] {$\Gamma$};
 \fill (1,5) circle(0.2);
 \fill (5,5) circle (0.2) node[anchor=west]{$(t_2(s_1),t_3(s_1))$} ;
\draw [thick,dashed] (1,1) .. controls (3,3) and (2,4) .. (5,5);
\end{tikzpicture}
\end{center}
\caption{ The deformation of the trajectory $\Gamma \rightarrow \Gamma_1+\Gamma_2 $  in the space of the independent variables $(t_1(s),t_2(s))$, where $s$ is the time parametised variable: $s_0<s<s_1$. With $\delta S=0$, we obtain the closure relation.}\label{x-curve}
\end{figure}
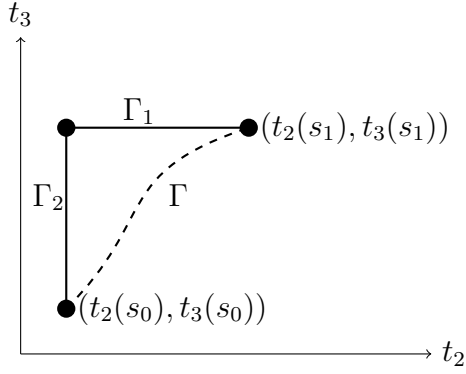

In this section, we need to consider the continuum limit of the remaining discrete variable $\mathsf N$. Again the naive limit on this variable would not satisfy our goal to create the hierarchy of the system. What we need is that we have to cook up the old set of semi-continuous variables $\{ \mathsf N,\tau_1,\tau_2,...\}$ to get a new set of fully continuous variables. In order to do so, we consider the factor    $(p+\sinh (\kappa))^{\mathsf N} e^ {\sinh(\kappa)\xi + \sum_{l=1} \mathrm \mu_l \tau_l}$ of the plane wave function form the previous section which can be rewritten
\begin{eqnarray}
(p+\sinh (\kappa))^{\mathsf N} e^{\sinh(\kappa)\xi+ \sum_{l=1} \mathrm  \mu_l \tau_l} &=& p^{\mathsf N} e^{\sinh(\kappa)\xi+ \sum_{l=1} \mathrm \mu_l \tau_l+\mathsf N\ln \left(1+\frac{\sinh (\kappa)}{p}\right)}\;.\label{CSOL}
\end{eqnarray}
We then expand the term ${\mathsf N}\ln \left(1+\frac{\sinh (\kappa)}{p}\right)$ with respect to the variable $p$,
\begin{eqnarray}\label{CSOL1}
{\mathsf N}\ln \left(1+\frac{\sinh (\kappa)}{p}\right) &=& \frac{{\mathsf N}}{p}\sinh (\kappa) - \frac{{\mathsf N}}{2p^2} \sinh^2 (\kappa)  + \frac{{\mathsf N}}{3p^3} \sinh^3 (\kappa) +...\;,
\end{eqnarray}
and also expand
\begin{subequations}
\begin{eqnarray}\label{CSOL2}
\mathrm \mu_1\tau_1 &=& -\frac{\tau_1}{p} \left (\frac{1}{1+\frac{\sinh(\kappa)}{p}}\right)
= -\frac{\tau_1}{p}+\frac{\tau_1}{p^2}\sinh(\kappa) -\frac{\tau_1}{p^3}\sinh^2(\kappa) +...\;,\\
\mathrm \mu_2\tau_2 &=&  -\frac{\tau _2}{2p^2}\left(\frac{1}{1+\frac {\sinh (\kappa)}{p}}\right)^2 
=-\frac{\tau_2}{2p^2} + \frac{\tau_2}{p^3}\sinh(\kappa) -  \frac{3\tau_2}{2p^4}\sinh^2(\kappa) +...\;,\\
\mathrm \mu_3\tau_3 &=&  -\frac{\tau _3}{3p^3}\left(\frac{1}{1+\frac {\sinh (\kappa)}{p}}\right)^3
= -\frac{\tau_3}{3p^3}+\frac{\tau_3}{p^4}\sinh(\kappa) -  \frac{2\tau_3}{p^5}\sinh^2(\kappa)+...\;,
\end{eqnarray}
\end{subequations}
which have been done only the first three terms. The remaining terms can be expanded in the same way. Hence, \eqref{CSOL} becomes 
\begin{eqnarray}\label{FLP}
(p+\sinh (\kappa))^{\mathsf N} e^{\sinh(\kappa)\xi+ \sum_{l=1} \mathrm  \mu_l \tau_l} &=&p^{\mathsf N}e^{\sinh{\kappa}\left(-\frac{\tau_1}{p}-\frac{\tau_2}{2p^2}-\frac{\tau_3}{3p^3}-...\right)} e^{\sinh(\kappa)\left(\xi +\frac{{\mathsf N}}{p}+\frac{\tau_1}{p^2}+...\right)} \nn\\
&&\times e^{\sinh^2{(\kappa)}\left(-\frac{{\mathsf N}}{2p^2}-\frac{\tau_1}{p^3}-\frac{3\tau_2}{2p^4}-...\right)} e^{\sinh^3(\kappa)\left(\frac{{\mathsf N}}{3p^3}+\frac{\tau_1}{p^4}+\frac{2\tau_2}{p^5}+...\right)} \nn\\
&&\times e^{\sinh^4{(\kappa)}\left(-\frac{{\mathsf N}}{4p^4} -\frac{\tau_1}{p^5}-\frac{5\tau_2}{2p^6}-...\right)}... \;.\nonumber
\end{eqnarray}
Then the plane wave function is in the form 
\begin{eqnarray}
\phi(t_1,t_2,t_3,...) = \left(1-\coth (\kappa) \sum_{i=1}^N \mathrm b_i \coth(t_1-X_i) \right) p^{\mathsf N} e^{-\sum_{i=1}^N{\frac{\tau_i}{ip^i} + \sum_{i=1}^N\sinh^i{(\kappa)}t_i}} \;,\label{CLPP}
\end{eqnarray}
where $\mathrm b_i =\mathrm b_i (t_1,t_2,t_3,...)$, ${\xi-\mathrm x_i} = {t_1-X_i}$, and
\begin{subequations}
\begin{eqnarray}\label{SYMB}
t_1&=&\xi +\frac{{\mathsf N}}{p}+\frac{\tau_1}{p^2}+\frac{\tau_2}{p^3} +\frac{\tau_3}{p^4}+\frac{\tau_4}{p^5}+...,\\
t_2&=&-\frac{{\mathsf N}}{2p^2}-\frac{\tau_1}{p^3}-\frac{3\tau_2}{2p^4}-\frac{2\tau_3}{p^5}-\frac{5\tau_4}{2p^6}...,\\
t_3&=&\frac{{\mathsf N}}{3p^3}+\frac{\tau_1}{p^4}+\frac{2\tau_2}{p^5}+\frac{10\tau_3}{3p^6}+\frac{5\tau_4}{p^7}+...\;.
\end{eqnarray}
\end{subequations}
\\
\\
\textbf{{Exact solution}}: We perform the expansion with respect to the variable $p$ on \eqref{EXACTSKEW2} yielding
\begin{eqnarray}\label{EXACTFULL}
\mathsf e^{{\boldsymbol Y}(t_1,t_2,...)}&=& e^{-\boldsymbol{\Lambda}(t_1-\xi)+\boldsymbol{\Lambda}^2t_2+...}   {\boldsymbol  Y(0,0)}e^{\boldsymbol{\Lambda}(t_1-\xi)-\boldsymbol{\Lambda}^2t_2+...} \nn\\
&&-(t_1-\xi)-2\boldsymbol{\Lambda}t_2 +...\;.
\end{eqnarray}
\\
\\
\textbf{Equations of motion}: To perform the continuum limit on the equations of motion, we first need to get the following relations
\begin{subequations}
\begin{eqnarray}
\frac{\partial {\mathsf x}_i}{\partial \tau_1}&=&\frac{\partial }{\partial \tau_1}\left(X_i -\frac{\mathsf N}{p}-\frac{\tau_1}{p^2}-\frac{\tau_2}{p^3}-\frac{\tau_3}{p^4}-...\right) \nn\\
&=&-\frac{1}{p^2} -\frac{1}{p^3}\frac{\partial X_i}{\partial t_2}+\frac{1}{p^4}\frac{\partial X_i}{\partial t_3}-\frac{1}{p^5}\frac{\partial X_i}{\partial t_4}+...\;,\\ \label{XDOT}
\frac{\partial {\mathsf x}_i}{\partial \tau_2}&=&\frac{\partial }{\partial \tau_2}\left(X_i -\frac{\mathsf N}{p}-\frac{\tau_1}{p^2}-\frac{\tau_2}{p^3}-\frac{\tau_3}{p^4}-...\right) \nn\\
&=&-\frac{1}{p^3} -\frac{3}{2p^4}\frac{\partial X_i}{\partial t_2}+\frac{2}{p^5}\frac{\partial X_i}{\partial t_3}-\frac{5}{2p^6}\frac{\partial X_i}{\partial t_4}+...\;, \\ \label{XSTAR}
\frac{\partial {\mathsf x}_i}{\partial \tau_3}&=&\frac{\partial }{\partial \tau_3}\left(X_i -\frac{\mathsf N}{p}-\frac{\tau_1}{p^2}-\frac{\tau_2}{p^3}-\frac{\tau_3}{p^4}-...\right) \nn\\
&=&-\frac{1}{p^4} -\frac{2}{p^5}\frac{\partial X_i}{\partial t_2}+\frac{10}{3p^6}\frac{\partial X_i}{\partial t_3}-\frac{5}{p^7}\frac{\partial X_i}{\partial t_4}+...\;, \label{XTRI}
\end{eqnarray}
\end{subequations} 
where $\frac{\partial X_i}{\partial t_1}=0$ is imposed.
We also find that
\begin{eqnarray}\label{NPM1}
 {\mathsf x}_i(\mathsf {N}\pm 1)&=&e^{\mp\frac{\partial}{p^2\partial t_2}\pm\frac{\partial}{p^3\partial t_3}\mp\frac{\partial}{p^4\partial t_5}\pm....}X_i\;\nn\\
 &=&X_i \mp \frac{1}{p} \mp \frac{1}{2p^2}\frac{\partial X_i}{\partial t_2}\pm  \frac{1}{3p^3}\frac {\partial X_i}{\partial t_3}
+\frac{1}{p^4}\left(\mp \frac{1}{4}\frac{\partial X_i}{\partial t_4}+\frac{1}{8}\frac{\partial^2 X_i}{\partial t_2^2} \right)\nn\\
&&+\frac {1}{p^5}\left(\pm \frac {1}{5}\frac{\partial X_i}{\partial t_5}-\frac {1}{6}\frac{\partial ^2 X_i}{\partial t_2\partial t_3}\right) +...\;.
\end{eqnarray}
We now ready to perform the continuum limit on the equations of motion.
\\
\\
\emph{The $\mathsf N-$flow}: Equation \eqref{DEQ0} gives
\begin{subequations}\label{HE1}
\begin{align}
\mathcal O(1/p^2)&:\;\;\;\;\frac{\partial^2 X_i}{\partial t_2^2}+ 8\sum_{j=1,j\neq i}^N\frac{\cosh (X_i-X_j)}{\sinh ^{3}(X_i-X_j)}= 0, \label{CONEQ01} \\
\mathcal O(1/p^3)&:\;\;\;\;\frac{\partial^2 X_i}{\partial t_2\partial t_3}
-6\sum_{j=1,j\neq i}^N\left( \frac{\partial X_i}{\partial t_2}+\frac{\partial X_j}{\partial t_2}\right) \frac{\cosh (X_i-X_j)}{\sinh ^3(X_i-X_j)} =0\;. \label{CONEQ02}
\end{align}
\end{subequations}
\\
\\
\emph{The $\tau_1-$flow}: Equation \eqref{DEQ1} gives
\begin{subequations}\label{HE2}
\begin{align}
\mathcal O(1/p^3)&:\frac{\partial^2 X_i}{\partial t_2^2}+ 8\sum_{j=1,j\neq i}^N\frac{\cosh (X_i-X_j)}{\sinh ^{3}(X_i-X_j)}= 0\;,\label{CONEQ11}\\
\mathcal O(1/p^4)&:
-\frac{1}{4}\frac{\partial X_i}{\partial t_2}\frac{\partial^2 X_i}{\partial t_2^2}-\frac{2}{3}\frac{\partial^2 X_i}{\partial t_2 \partial t_3}
+\sum_{j=1,j\neq i}^N\left(2 \frac{\partial X_i}{\partial t_2}+4\frac{\partial X_j}{\partial t_2}\right) \nn\\
&\times\frac{\cosh (X_i-X_j)}{\sinh^{3}(X_i-X_j)} =0. \label{CONEQ12}
\end{align}
Equation \eqref{CONEQ12} can be further simplified by using the\eqref{CONEQ11} resulting to
\begin{eqnarray}
&&\frac{\partial^2 X_i}{\partial t_2\partial t_3} -6\sum_{j=1,j\neq i}^N\left( \frac{\partial X_i}{\partial t_2}+\frac{\partial X_j}{\partial t_2}\right) \frac{\cosh (X_i-X_j)}{\sinh ^{3}(X_i-X_j)} =0 \;.\label{CONEQ13}
\end{eqnarray}
\end{subequations}
\\
\\
\emph{The $\tau_2-$flow}: Equation \eqref{DEQ2} gives
\begin{subequations}\label{HE3}
\begin{align}
\mathcal O(1/p^4)&:\;\;\;\frac{\partial^2 X_i}{\partial t_2^2}+ 8\sum_{j=1,j\neq i}^N\frac{\cosh (X_i-X_j)}{\sinh ^{3}(X_i-X_j)}= 0\;,\label{CONEQ21}\\
\mathcal O(1/p^5)&:
\frac{\partial X_i}{\partial t_2}\frac{\partial^2 X_i}{\partial t_2^2}+\frac{5}{3}\frac{\partial^2 X_i}{\partial t_2 \partial t_3} = \sum_{j=1,j\neq i}^N\left(2\frac{\partial X_i}{\partial t_2}+10\frac{\partial X_j}{\partial t_2}\right) \frac{\cosh (X_i-X_j)}{\sinh ^{3}(X_i-X_j)} \;.\label{CONEQ22}
\end{align}
Again, equation \eqref{CONEQ22} can be simplified by using the equation \eqref{CONEQ21} yielding
\begin{eqnarray}
&&\frac{\partial^2 X_i}{\partial t_2\partial t_3}-6\sum_{j=1,j\neq i}^N\left( \frac{\partial X_i}{\partial t_2}+\frac{\partial X_j}{\partial t_2}\right) \frac{\cosh (X_i-X_j)}{\sinh ^{3}(X_i-X_j)} =0\;. \label{CONEQ23}
\end{eqnarray}
\end{subequations} 
In \eqref{HE1}, \eqref{HE2} and \eqref{HE3}, only the second and third flows in the heirarchy are considered. We observe that every flows of the semi-continuous time case produce the same fully continuous-time hierarchy of the hyperbolic Calogero-Moser system.
\\
\\
\textbf{{Lagrangians}}:
The Lagrangians for the first two flows in the fully continuous-time hierarchy of the hyperbolic Calogero-Moser system are given by
\begin{eqnarray}\label{eqc}
\mathscr L_{(t_2)}&=&\sum\limits_{i=1}^N\frac{1}{2}\left( \frac{\partial X_i}{\partial t_2}\right)^{2} + 2\sum\limits_{i \ne j}^N \csch^2 (X_i-X_j) \;, \label{La1}\\
\mathscr L_{(t_3)}&=&\sum\limits_{i=1}^N\left(\frac{\partial X_i}{\partial t_2}\frac{\partial X_i}{\partial t_3}+\frac{1}{4}\left( \frac{\partial X_i}{\partial t_2}\right)^{3}\right) -3\sum\limits_{i \ne j}^N\frac{\partial X_i}{\partial t_2}\csch^{2}(X_i-X_j)\;.\label{La2}
\end{eqnarray}
and their Euler-Lagrange equations are given by
\begin{eqnarray}\label{EL}
&&\frac{\partial \mathscr L_{(t_2)}}{\partial X_i}-\frac{\partial}{\partial t_2}\left( \frac{\partial 
 \mathscr L_{(t_2)}}{\partial(\frac{\partial X_i}{\partial t_2})}\right)=0\;, \\
 &&\frac{\partial \mathscr L_{(t_3)}}{\partial X_i}-\frac{\partial}{\partial t_3}\left( \frac{\partial 
\mathscr L_{(t_3)}}{\partial(\frac{\partial X_i}{\partial t_3})}\right)=0\;.
\end{eqnarray}
Obviously, these Euler-Lagrange produce the equations of motion given in \eqref{HE1}, \eqref{HE2} and \eqref{HE3}.
\\
\\
If we are restrict our interest for only the first two flows the action of the system is
\begin{eqnarray}
S[X] &=&\int_{\Gamma}\left( \mathscr L_{(t_2)}dt_2 + \mathscr L_{(t_3)}dt_3\right)=\int_{s_0}^{s_1}\left( \mathscr L_{(t_2)}\frac{dt_2}{ds} + \mathscr L_{(t_3)}\frac{dt_3}{ds}\right)ds \;
\end{eqnarray}
for an arbitrary dashed curve $\Gamma$ on the space of independent variables shown in figure \ref{x-curve}.
Performing the local variation on the space of independent variables with condition $\delta S=0$, we obtain \cite{Sikarin1, Sikarin2}
\begin{equation}\label{CR}
\frac{\partial \mathscr L_{(t_2)} }{\partial t_3}=\frac{\partial \mathscr L_{(t_3)} }{\partial t_2}\;,
\end{equation}
which is nothing but the continuous-time version of the closure relation between $\mathscr L_{(t_2)}$ and $\mathscr L_{(t_3)}$. 
\\
\\
\emph{Proof}: Here we will prove the closure relation \eqref{CR} through the direct computation. We first compute 
\begin{eqnarray}\label{fullproof1}
\frac{\partial \mathscr L_{(t_2)}}{\partial t_3}&=& \sum_{i=1}^N
\frac{\partial X_i}{\partial t_2}\frac{\partial^2 X_i}{\partial t_2\partial t_3}-4\sum\limits_{i \ne j}^N\left(\frac{\partial X_i}{\partial t_3}-\frac{\partial X_j}{\partial t_3}\right)\frac{\cosh(X_i-X_j)}{\sinh^3(X_i-X_j)}\;.
\end{eqnarray}
Using \eqref{CONEQ02}, \eqref{CONEQ13}, or \eqref{CONEQ23}, \eqref{fullproof1} is simplified to 
\begin{eqnarray}\label{fullproof11}
\frac{\partial \mathscr L_{(t_2)}}{\partial t_3} &=& \sum\limits_{i \ne j}^N\left(6\left(\frac{\partial X_i}{\partial t_2}\right)^2 -8\frac{\partial X_i}{\partial t_3}\right)\frac{\cosh(X_i-X_j)}{\sinh^3(X_i-X_j)}\;,\;\;\;\;
\end{eqnarray}
and we also compute
\begin{eqnarray}\label{fullproof2}
\frac{\partial \mathscr L_{(t_3)}}{\partial t_2}&=& \sum\limits_{i \ne j}^N\left(6\left(\frac{\partial X_i}{\partial t_2}\right)^2 -8\frac{\partial X_i}{\partial t_3}\right)\frac{\cosh(X_i-X_j)}{\sinh^3(X_i-X_j)} \;\;\;\;\nn\\
&&-3\sum\limits_{i \ne j}^N\frac{\partial^2 X_i}{\partial t_2^2}\csch^2(X_i-X_j)\;.
\end{eqnarray}
Using \eqref{fullproof11} and \eqref{fullproof2}, we find that
\begin{equation}\label{CR2}
\frac{\partial \mathscr L_{(t_2)} }{\partial t_3}-\frac{\partial \mathscr L_{(t_3)} }{\partial t_2}=3\sum\limits_{i \ne j}^N\frac{\partial^2 X_i}{\partial t_2^2}\csch^2(X_i-X_j)\;.
\end{equation}
Next, we consider the term on the right-hand-side of \eqref{CR2} and rewrite in the form
\begin{eqnarray}\label{FF}
\frac{1}{8}\sum\limits_{i \ne j}^N\frac{\partial^2 X_i}{\partial t_2^2} \csch^2(X_i-X_j) &=&-\frac{1}{8}\sum\limits_{i \ne j}^N \csch^2(X_i-X_j) 8\sum\limits_{k \ne i}^N\frac{\cosh(X_i-X_k)}{\sinh^3(X_i-X_k)}\nn\\
&=&-\sum\limits_{i \ne j }^N\sum\limits_{ i \ne k }^N\frac{\csch^2(X_i-X_j) \cosh(X_i-X_k)}{ \sinh^3(X_i-X_k)}\nn\\
&=&-\sum\limits_{i \ne j}^N\frac {\cosh(X_i-X_j)}{\sinh^5(X_i-X_j)}\nn\\
&&-\sum\limits_{i \ne j \ne k }^N\frac{\csch^2(X_i-X_j) \cosh(X_i-X_k)}{\sinh^3(X_i-X_k)}\;.
\end{eqnarray}
The first term of \eqref{FF} is antisymmetric, so vanishes. Using the hyperbolic function identity, we rewrite the second term of \eqref{FF} as
\begin{eqnarray}\label{FF2}
&&\sum\limits_{i \ne j \ne k }^N\frac{\csch^2(X_i-X_j) \cosh(X_i-X_k)}{ \sinh^3(X_i-X_k)} \;\nn\\
&&\;\;\;\;\;\;\;\;\;\;\;\;=\sum\limits_{i \ne j \ne k }^N\left( \coth^2 (X_i-X_j) -1 \right)\left(\coth^2(X_i-X_k)-1\right)\coth(X_i-X_k) \;.
\end{eqnarray}
To simplify the right-hand-side of \eqref{FF2}, we need the identity 
\begin{eqnarray}\label{IDEN}
\coth(X_i-X_j)\coth(X_i-X_k)=\coth(X_k-X_j)\left(\coth(X_i-X_k)-\coth(X_i-X_j)\right) +1\;. 
\end{eqnarray}
We now define 
\begin{eqnarray}
A \equiv \coth(X_i-X_j) ,\;\;\;\; 
B \equiv \coth(X_i-X_k) ,\;\;\;\; 
C \equiv \coth(X_k-X_j)\;,\;\;\; \nonumber
\end{eqnarray}
then \eqref{IDEN} becomes 
\begin{eqnarray}\label{IDEN1}
AB &=&CB - CA +1\;. 
\end{eqnarray}
\eqref{FF2} can be rewritten in terms of $A$, $B$ and $C$ as follows
\begin{eqnarray}\label{FF3}
\sum \limits_{i \ne j \ne k }^N\frac{\csch^2(X_i-X_j) \cosh(X_i-X_k)}{ \sinh^3(X_i-X_k)} &=&
A^2B^3-A^2B-B^3+B \;.
\end{eqnarray}
The third and fourth terms of \eqref{FF3} are antisymmetric, hence vanish. Using \eqref{IDEN1}, we may have \eqref{FF3} in the form
\begin{eqnarray}\label{FF4}
\sum \limits_{i \ne j \ne k }^N\frac{\csch^2(X_i-X_j) \cosh(X_i-X_k)}{ \sinh^3(X_i-X_k)} &=&AB^2(CB-CA+1)-A^2B \nn\\
&=&AB^3C-A^2B^2C+AB^2-A^2B.\;
\end{eqnarray}
The first and second terms of \eqref{FF4} again vanish due to the antisymmetric property. The third and fourth terms of \eqref{FF4} cancel each other. Then we can conclude that
\begin{equation}\label{CR3}
\frac{\partial \mathscr L_{(t_2)} }{\partial t_3}-\frac{\partial \mathscr L_{(t_3)} }{\partial t_2}=0\;.
\end{equation}
Again, the closure relation guarantees that of the action is invariant under local deformation of the curve on the space of independent variables, see figure \ref{x-curve}.
\section{Summary}
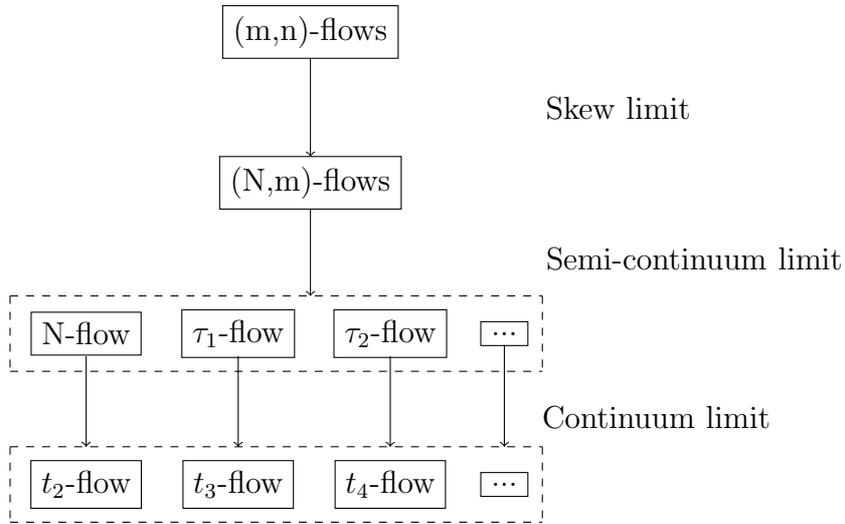
\begin{figure}[h]
\begin{center}
\begin{tikzpicture}
 \node (nm)   at (1.95,4)  [draw] {(m,n)-flows};
                      \draw[->] (1.95,3.65) -- (1.95,2.35);
                       \draw (6,3) node {Skew limit};
    \node (Nm)   at (1.95,2)  [draw] {(N,m)-flows};
            \draw[->] (1.95,1.65) -- (1.95,0.5); 	 
           \draw (7,1) node {Semi-continuum limit};
           \draw [dashed] (-2,0.5) -- (5,0.5);
           \draw [dashed] (-2,0.5) -- (-2,-0.5); 
            \draw [dashed] (-2,-0.5) -- (5,-0.5); 
            \draw [dashed] (5,0.5) -- (5,-0.5);  
    \node (Nf)    at (-1,0) [draw] {N-flow};
  \node (tau1)  at (1,0)  [draw] {$\tau_1$-flow};
  \node (tau2) at (3,0)  [draw] {$\tau_2$-flow};
  \node (tau.) at (4.5,0)  [draw] {...};
            \draw[->] (-1,-0.3) -- (-1,-1.5);
            \draw[->] (1,-0.3) -- (1,-1.5);
            \draw[->] (3,-0.3) -- (3,-1.5);
             \draw[->] (4.5,-0.15) -- (4.5,-1.5);
  \draw (6.5,-1.1) node {Continuum limit}; 
   \node (t2)    at (-1,-2) [draw] {$t_2$-flow};
  \node (t3)  at (1,-2)  [draw] {$t_3$-flow};
  \node (t4) at (3,-2)  [draw] {$t_4$-flow};
  \node (t.) at (4.5,-2)  [draw] {...}; 
          \draw [dashed] (-2,-1.5) -- (5,-1.5);
          \draw [dashed] (-2,-1.5) -- (-2,-2.5); 
          \draw [dashed] (-2,-2.5) -- (5,-2.5); 
          \draw [dashed] (5,-1.5) -- (5,-2.5); 
 \end{tikzpicture}
\end{center}
\caption{ The flow chat shows the discrete-time, semi-continuous, and fully-continuous hierarchies.}\label{xx-curve}
\end{figure}

In this present work, we report the structure of Lagrangian 1-form of the hyperbolic Calogero-Moser system. Employing the pole-reduction method, the two discrete-time hyperbolic Calogero-Moser systems are obtained from the semi-discrete KP equation. The key relation called the closure relation is established through the connection between the discrete-time Lagrangian and the temporal Lax matrix. The skew limit, but with a little modification, is performed on the discrete-time system resulting to the hierarchy of the semi-continuous time hyperbolic Calogero-Moser system. The full limit is then performed to obtain the hierarchy of the continuous-time hyperbolic Calogero-Moser system. We find that each flow in the hierarchy of the semi-continuous time hyperbolic Calogero-Moser system produces the same continuous-time hierarchy of the hyperbolic Calogero-Moser system and show in figure \ref{xx-curve}.
\\
We knew that apart from the hyperbolic Calogero-Moser system there are the hyperbolic Ruijsenaar-Schneider system and the hyperbolic Goldfish system. In appendix \ref{C.}, we manage to report their Lagrangian structures. Furthermore, we also knew that there is the elliptic Calogero-Moser type systems. The connection between temporal elliptic Lax matrix and Lagrangian was established in \cite{Sikarin1} for the elliptic Calogero-Moser system. This means that we could directly get the closure relation from the compatibility of the temporal Lax matrices. For the elliptic Ruijsenaar-Schneider system and the elliptic Goldfish system, the connection between the temporal Lax matrix and the Lagrangian do not exist, forcing us to perform the explicit computation in order to get the closure relation. The present difficulty is that the exact solutions for these discrete-time elliptic systems are not yet obtained. One more remark is about the quantum version of the Lagrangian 1-form in the view of the Feynman Path integral. This task is worth to pursue in forthcoming publications.
\appendix
\section{The Lax Pair}\label{A.}
\numberwithin{equation}{section}
We reconsider a semi-discrete KP equation
 \begin{eqnarray}
\partial_ \xi(\wh{u}-\wt{u})&=& (p-q+\wh{u}-\wt{u})(u+{\widehat{\widetilde u}}-\wh{u}-\wt{u})\ .\label{SemiKPP1}
\end{eqnarray}
We find that  \eqref{SemiKPP1} is a result of compatibility between
\begin{subequations}\label{SLPP} 
\begin{eqnarray}
\wt{ \phi}&=&\phi_ \xi +(p+u-\wt{u})\phi\,,\label{SLPP1}\\
\wh{ \phi}&=&\phi_ \xi +(q+u-\wh{u})\phi\,.\label{SLPP2}
\end{eqnarray}
\end{subequations}
We find that $\phi$ and $u$ take the form
\begin{subequations}
\begin{eqnarray}
u &=&\sum_{l=1}^N\coth{(\xi-x_i)}\;,\label{uu} \\
\phi &=& \left(1-\coth (\kappa) \sum_{i=1}^N b_i \coth (\xi -x_i) \right)(p+\sinh (\kappa))^n (q+\sinh (\kappa))^m e^{\sinh(\kappa)\xi}\;.\label{SLPP}
\end{eqnarray}
\end{subequations}
From \eqref{uu}, we find
\begin{eqnarray}
\wt{\phi} &=& \left(1-\coth (\kappa) \sum_{i=1}^N \wt{b_i} \coth (\xi -\wt{x}_i) \right)(p+\sinh (\kappa))^{n+1}(q+\sinh (\kappa))^m e^{\sinh(\kappa)\xi}\;,\label{SLPPP}
\end{eqnarray}
and  from \eqref{SLPP}, we find
\begin{eqnarray}\label{SLPPP1}
\phi_\xi  &=& -\coth (\kappa) \sum_{i=1}^N b_i  \left( 1-\coth^2 (\xi -x_i) \right)(p+\sinh (\kappa))^n (q+\sinh (\kappa))^m e^{\sinh(\kappa)\xi}\nn\\
&&+\left(1-\coth (\kappa) \sum_{i=1}^Nb_i  \coth (\xi -x_i) \right)\sinh (\kappa)(p+\sinh (\kappa))^n (q+\sinh (\kappa))^m e^{\sinh(\kappa)\xi} \;.\nn\\
\end{eqnarray}
Using above relations, \eqref{SLPP1} gives
\begin{eqnarray}\label{SLPPP2}
&&\left(1 -\coth (\kappa) \sum_{l=1}^N \wt{b_l}( 1-\coth(\xi -\wt{x}_i))\right)(p+\sinh (\kappa)) = \nn\\
&&-\coth (\kappa)\left( \sum_{l=1}^N(1-\coth ^2(\xi -x_i))\right) +\left(1-\coth (\kappa)\sum_{l=1}^N b_l \coth(\xi -x_i)\right) \sinh (\kappa)\nn\\
&&+\left(p+\sum_{l=1}^N \coth (\xi -x_l)-\sum_{l=1}^N \coth(\xi -\wt{x}_l)\right)\left(1-\coth (\kappa)\sum_{m=1}^N b_m \coth (\xi -x_m)\right)\;.\nn\\ 
\end{eqnarray}
Equating coefficients of $\coth (\kappa)\coth(\xi -\wt{x}_l)$ in \eqref{SLPP2}, we have 
 \begin{eqnarray}
-\left(p+\sinh (\kappa)\right)\wt{b_l}&=& -\frac{1}{\coth(\kappa)} - \sum_{m=1}^Nb_m\coth (x_m-\wt{x}_l)\;.\label{SemiKPP}
\end{eqnarray}
Rewriting above equation in the matrix form, we get
\begin{eqnarray}\label{LP} 
(p+\sinh (\kappa))\wt{\boldsymbol b} &=&\tanh (\kappa)\boldsymbol E +\boldsymbol M \boldsymbol b\;,
\end{eqnarray}
where $\boldsymbol b=(b_1,b_2,...,b_N)^T$, $\boldsymbol E=(1,1,...,1)^T$ and 
\begin{eqnarray}
\boldsymbol M &=& -\sum_{i,j=1}^N\coth (\wt{x}_i-x_j)E_{ij}\; .\,\label{MHCMM}
\end{eqnarray}
Equating coefficient of $\coth (\kappa)\coth(\xi -x_l)$ in \eqref{SLPP2}, we have 
\begin{eqnarray}
(p+\sinh (\kappa))b_l & =& \sum_{m=1}^N b_l\coth (x_l-\wt{x}_m) -  \sum_{m=1}^N b_l\coth (x_l-x_m)\nn\\
&&-\sum_{m=1}^N b_m \coth (x_l-x_m)+\frac{1}{\coth(\kappa)} \;,\label{SemiKPPP}
\end{eqnarray}
and its matrix form is
\begin{eqnarray}\label{LPP2} 
(p+\sinh (\kappa))\boldsymbol b &=&\tanh (\kappa)\boldsymbol E +\boldsymbol L \boldsymbol b\;,
\end{eqnarray}
where
\begin{eqnarray}\label{LHCMM}
\boldsymbol L&=&\sum_{i,j=1}^N \left(\coth{(x_i-\wt{x}_j)}- \coth{(x_i-x_j)}\right)E_{ii} -\sum_{j \ne i}^N\coth (x_i-x_j)E_{ij}\;.
\end{eqnarray}
\\
The compatibility between \eqref{LP} and \eqref{LP2} gives
\begin{eqnarray}\label{COMMLLM}
\wt{\boldsymbol L}\boldsymbol M-\boldsymbol M \boldsymbol L=\tanh(\kappa)\left(\boldsymbol M \boldsymbol E+\boldsymbol E\boldsymbol L-\wt{\boldsymbol L}\boldsymbol E-\boldsymbol E\boldsymbol M\right) \;.
\end{eqnarray}
We find that on the left hand side of \eqref{COMMLLM} gives us
\begin{eqnarray} \label{LMMLL}
\wt{\boldsymbol L} \boldsymbol M &=& \boldsymbol M \boldsymbol L \;,
\end{eqnarray}
and on the right hand side of \eqref{COMMLLM} gives us
\begin{eqnarray}\label{LMMLLM}
\left(\wt{\boldsymbol L}-\boldsymbol M \right)\boldsymbol E&=&\boldsymbol E \left(\boldsymbol L -\boldsymbol M \right)\;.
\end{eqnarray}
\section{The exact solution}\label{B.}
In this section, we will solve for $x_i(n,m)$. We start to rewrite matrix $\boldsymbol L$ and matrix $\boldsymbol M$ \cite{Sikarin1} as
\begin{subequations}\label{BB}
\begin{eqnarray}
e^{\wt{\boldsymbol X}}  \boldsymbol{M}-\boldsymbol{M} e^{\boldsymbol X}&=&-\boldsymbol E\;,\label{MX}\\
e^{\boldsymbol X}  \boldsymbol{L}-\boldsymbol{L} e^{\boldsymbol X}&=&\boldsymbol I-\boldsymbol E\;,\label{LX}
\end{eqnarray}
where $\boldsymbol X=\sum_{i=1}^Nx_iE_{ii}$ is the diagonal matrix of the particle positions. We also have
\begin{eqnarray}
(\wt{\boldsymbol{L}}-\boldsymbol{M})\boldsymbol E&=&0\;,\label{L0M}\\
\boldsymbol E(\boldsymbol L- \boldsymbol M )&=&0\;.\label{L0MM}
\end{eqnarray}
\end{subequations}
Using $\boldsymbol{M}=\wt{\boldsymbol U}\boldsymbol U^{-1}$ and $\boldsymbol L=\boldsymbol U \bLam \boldsymbol U^{-1}$, where $\boldsymbol U=\boldsymbol U(n,m) $ is an invertible matrix used to diagonalise the matrix $\boldsymbol L$, \eqref{BB} becomes
\begin{eqnarray}
e^{\wt{\boldsymbol Y}} - e^{\boldsymbol Y}&=&-\wt{\boldsymbol U}^{-1}\boldsymbol E \boldsymbol U\;,\label{EMX}\\
e^{\boldsymbol Y}  \boldsymbol{\bLam}- \boldsymbol{\bLam} e^{\boldsymbol Y}&=&\boldsymbol U^{-1}(\boldsymbol I-\boldsymbol E )\boldsymbol U \;,\label{LX}\\
\wt{\boldsymbol U}^{-1}\boldsymbol E&=&\bLam^{-1}\boldsymbol U\boldsymbol E\;,\label{UE}\\
\boldsymbol E\wt{\boldsymbol U}&=&\boldsymbol E\boldsymbol U\bLam\;,\label{EUU}
\end{eqnarray}
where
\begin{eqnarray}
e^{\boldsymbol Y(n,m)} &=&\boldsymbol U^{-1} e^{\boldsymbol X(n,m)} \boldsymbol U \;.\label{LX2}
\end{eqnarray}
Multiplying \eqref{EMX} by $\boldsymbol{\bLam}$ on the right hand side, we get
\begin{eqnarray}
e^{\wt{\boldsymbol Y}} \boldsymbol{\bLam}- e^{\boldsymbol Y}\boldsymbol{\bLam}&=&-\wt{\boldsymbol U}^{-1}\boldsymbol E \boldsymbol U\boldsymbol{\bLam}\;.\label{Dq}
\end{eqnarray}
Taking ``$\;\;\wt{}\;\;$" on \eqref{LX}, we obtain
\begin{eqnarray}
e^{\wt{\boldsymbol Y}}  \boldsymbol{\bLam} -\boldsymbol{\bLam}  e^{\wt{\boldsymbol Y}} &=& \boldsymbol I-\wt{\boldsymbol U}^{-1}\boldsymbol E \wt{\boldsymbol U }\;,\nn\\
e^{\wt{\boldsymbol Y}}  \boldsymbol{\bLam} -\boldsymbol{\bLam} e^{\wt{\boldsymbol Y}} - \boldsymbol I &=& -\wt{\boldsymbol U}^{-1}\boldsymbol E \boldsymbol U \boldsymbol \Lambda\;,\label{EMXX}
\end{eqnarray}
Equating \eqref{Dq} with \eqref{EMXX}, we obtain
\begin{eqnarray}
 e^{\wt{\boldsymbol Y}} &=& \boldsymbol{\bLam}^{-1} e^{\boldsymbol Y} \boldsymbol{\bLam} -\boldsymbol{\bLam}^{-1}  \;.\label{EEMXX}
\end{eqnarray}
Shifting on \eqref{EEMXX}, we find that
\begin{eqnarray}\label{BS}
e^{\wt{\wt{\boldsymbol Y}}} &=& \boldsymbol{\bLam}^{-1} e^{\wt{\boldsymbol Y}} \boldsymbol{\bLam} -\boldsymbol{\bLam}^{-1}\nn\\
 &=&\boldsymbol{\bLam}^{-1} \left(\boldsymbol{\bLam}^{-1}e^{\boldsymbol Y} \boldsymbol{\bLam} -\boldsymbol{\bLam}^{-1}\right)\boldsymbol{\bLam} - \boldsymbol{\bLam}^{-1} \;\nn\\
&=&\boldsymbol{\bLam}^{-2}e^{\boldsymbol Y} \boldsymbol{\bLam}^{2} -2\boldsymbol{\bLam}^{-1}\;.
\end{eqnarray}
Repeating the process in \eqref{BS} $a$ times, we get
\begin{eqnarray}\label{Yn}
e^{\boldsymbol{Y}(n+a,m)}=(p\boldsymbol{I}+\bLam)^{-a} e^{\boldsymbol Y(n,m)}(p\boldsymbol{I}+\bLam)^{a} -\frac{a}{p\boldsymbol{I}+\bLam} \;.
\end{eqnarray}
Similarly, we find the solution in $m-$direction as
\begin{eqnarray}\label{Ym}
e^{\boldsymbol{Y}(n,m+b)}=(q\boldsymbol{I}+\bLam)^{-b} e^{\boldsymbol Y(n,m)}(q\boldsymbol{I}+\bLam)^{b} -\frac{b}{q\boldsymbol{I}+\bLam} \;,
\end{eqnarray}
where we replace $\bLam \rightarrow  p\boldsymbol{I}+\bLam$ in \eqref{Yn} and $\bLam \rightarrow  q\boldsymbol{I}+\bLam$ in \eqref{Ym}, since $ \boldsymbol{L}$ and $ \boldsymbol{K}$ differ by $(p-q)\boldsymbol I$ \cite{Sikarin1}.
Finally, the combination of both discrete-time directions gives
\begin{eqnarray}\label{Ynm}
e^{\boldsymbol{Y}(n+a,m+b)}&=&(p\boldsymbol{I}+\bLam)^{-a}(q\boldsymbol{I}+\bLam)^{-b} e^{\boldsymbol Y(n,m)}(q\boldsymbol{I}+\bLam)^{b}(p\boldsymbol{I}+\bLam)^a\nn\\
&&-\frac{a}{p\boldsymbol{I}+\bLam} -\frac{b}{q\boldsymbol{I}+\bLam}\;.
\end{eqnarray}

\section{The Lagrangians for hyperbolic Ruijsenaaars-Schneider (HRS) and the hyperbolic Goldfish (HGF) system} \label{C.}
In this section, we report some preliminary results on the Lagrangian structure of the HRS and HGF systems.
\\
\\
\textbf{\emph{The discrete-time case}}.
\\
\\
The Ansatz forms of the Lax matrices $\boldsymbol L_\kappa$ and $\boldsymbol M_\kappa$ are given by
\\
\\
\textbf{\emph{HRS}}:
\begin{subequations}
\begin{eqnarray}\label{coN}
\boldsymbol L_\kappa &=&\sum_{i,j=1}^N{h_ih_j}\left(\coth (\kappa)+\coth (x_i-x_j+\lambda)\right)E_{ij}\;,
\end{eqnarray}
 \begin{eqnarray}\label{conh2}
\boldsymbol M_\kappa &=&\sum_{i,j=1}^N{\wt{h}_ih_j}\left(\coth (\kappa)+\coth (\wt{x}_i-x_j+\lambda)\right)E_{ij}\;,
 \end{eqnarray}
\end{subequations}
\textbf{\emph{HGF}}:
\begin{subequations}
\begin{eqnarray}\label{coNM}
\boldsymbol L_\kappa &=&\sum_{i,j=1}^N{h_ih_j}\left(\coth (\kappa)+\coth (x_i-x_j)\right)E_{ij}\;,
\end{eqnarray}
\begin{eqnarray}\label{conh3}
\boldsymbol M_\kappa &=&\sum_{i,j=1}^N{\wt{h}_ih_j}\left(\coth (\kappa)+\coth (\wt{x}_i-x_j)\right)E_{ij}\;.
\end{eqnarray}
\end{subequations}
From these Lax matrices, we can find the discrete-time equations of motion of the systems \cite{Sikarin2, Umpon} 
\\
\\
\textbf{\emph{HRS}}:
\begin{eqnarray}\label{eqd2}
\prod\limits_{\mathop {j = 1}}^N \frac{\sinh (x_i-\wt{x}_j) \sinh (x_i-\hypotilde 0 {x}_j+\lambda)}{\sinh (x_i-\hypotilde 0 {x}_j) \sinh (x_i-\wt {x}_j-\lambda)} &=&
\prod\limits_{\mathop {j = 1},j \ne i}^N\frac{\sinh(x_i-x_j+\lambda)}{\sinh(x_i-x_j-\lambda)}
\;,
\end{eqnarray}
\textbf{\emph{HGF}}:
\begin{eqnarray}\label{eqmotion11}
\prod\limits_{\mathop {j = 1}}^N\frac{\sinh (x_i-\wt{x}_j)}{\sinh (x_i-\hypotilde 0 {x}_j)}&=&-1\;,
\end{eqnarray}
and their discrete-time Lagrangians are given by
\\
\\
\textbf{\emph{HRS}}:
\begin{eqnarray}
\mathscr{L}(\boldsymbol x,\wt{\boldsymbol x}) &=& \sum_{i,j=1}^N \left(\int_{0}^{(x_i-\wt{x}_j)}{ \ln |{\sinh \theta} |} d\theta - \int_{0}^{(x_i-\wt{x}_j-\lambda)} \ln |\sinh\theta | d\theta\right) \nn\\ 
&&-\sum_{j \ne i}^N \int_{0}^{(x_i-x_j+\lambda)}  \ln  |  \sinh \theta | d\theta + p(\Xi -\wh{\Xi })\; ,
\end{eqnarray}
\textbf{\emph{HGF}}:
\begin{eqnarray}
\mathscr{L}(\boldsymbol x,\wt{\boldsymbol x})&=&\sum_{i,j=1}^N \left( \int_{0}^{(x_i-\wt{x}_j)}\ln |\sinh \theta |\mathrm{d}\theta\right)+p(\Xi -\wh{\Xi })\;.
\end{eqnarray}
\textbf{\emph{The semi-continuous time case}}.
\\
\\
We proceed the same method provided in \cite{Sikarin2,Umpon} to obtain
\\
\\
\textbf{\emph{HRS}}:
\begin{eqnarray}\label{eq12}
&&\sum_{j \ne i}^N \ln\left|\frac{\sinh(\mathrm  x_i-\mathrm  x_j+\lambda)}{\sinh(\mathrm  x_i-\mathrm  x_j-\lambda)}\right| = \sum_{j=1}^N \ln\left|\frac{\sinh (\mathrm  x_i-{\bar{\mathrm  x}}_j)\sinh (\mathrm x_i-\underline  {\mathrm {x}}_j+\lambda)}{\sinh(\mathrm x_i-\underline  {\mathrm {x}}_j)\sinh (\mathrm  x_i-{\bar{\mathrm  x}}_j+\lambda)}\right| \;,\label{x11}\\
&&\sum_{j=1}^N\left(\frac{\partial {\bar {\mathrm x}_j}}{\partial \tau}\left( \coth(\mathrm x_i-{\bar{\mathrm  x}}_j)- \coth(\mathrm x_i-{\bar{\mathrm  x}}_j-\lambda)\right)\right.\nn\\
&&\;\;\;\;\;\;\;\;\;\;\;\;\;\;\;\;\;\;\;\;\;\;\;\;\;\;\left.+\frac{\partial \underline {\mathrm x}_j}{\partial \tau}\left(\coth(\mathrm x_i-\underline {\mathrm x}_j+\lambda)-\coth(\mathrm x_i-\underline {\mathrm x}_j)\right) \right)=0 \;.\label{x12}
\end{eqnarray}
\textbf{\emph{HGF}}:
\begin{eqnarray}\label{eq12}
&&\sum_{j=1}^N\ln\left|\frac{\sinh ({\bar{\mathrm  x}}_j-\mathrm  x_i)}{\sinh (\mathrm x_i-\underline  {\mathrm {x}}_j)}\right| = 0 \;,\label{x11}\\
&&\sum_{j=1}^N\left(\frac{\partial {\bar {\mathrm x}_j}}{\partial \tau}\coth({\bar{\mathrm  x}}_j-\mathrm x_i)+\frac{\partial \underline {\mathrm x}_j}{\partial \tau}\coth(\mathrm x_i-\underline {\mathrm x}_j)\right)=0\;.\label{x12}
\end{eqnarray}
\textbf{\emph{The continuous-time case}}.
\\
\\
We proceed the same method provided in \cite{Sikarin2, Umpon} to get the system in continuous-time case. Here we give the equations of motion only for the first two flows.
\\
\\
\textbf{\emph{HRS}}:
\begin{subequations}
\begin{eqnarray}\label{EQMOTIONRS1}
\frac{\partial^2 X_i}{\partial t_1^2}/\frac{\partial X_i}{\partial t_1} &=& \sum_{j=1,j\neq i}^N \frac{\partial X_j}{\partial t_2}\left( \coth (X_i-X_j+\lambda) +\coth (X_i-X_j-\lambda)\right.\;\;\;\;\;\nn\\
&&\left.-2\coth (X_i-X_j)\right),\;
\end{eqnarray}
\begin{eqnarray}\label{EQMOTIONRS1}
\frac{\partial^2 X_i}{\partial t_1\partial t_2} &=& -\sum_{j=1,j\neq i}^N\left(\frac{\partial X_j}{\partial t_2}\left[ \coth (X_i-X_j+\lambda)+\coth (X_i-X_j-\lambda)\right.\right. \nn\\
&&\;\;\;\left.\left.-2\coth (X_i-X_j)\right] -\frac{1}{2} \frac{\partial X_i}{\partial t_1}\frac{\partial X_j}{\partial t_1} \left[\coth (X_i-X_j-\lambda) \right. \right.\nn\\
&&\left.\phantom{\frac{1}{2}}\left.+\coth (X_i-X_j+\lambda) \right]\right) .\;
\end{eqnarray}
\end{subequations}
\\
\textbf{\emph{HGF}}:
\begin{eqnarray}\label{EQMOTIONRS1}
\frac{\partial^2 X_i}{\partial t_1^2}&=& 2\sum_{j=1,j\neq i}^N\frac{\partial X_i}{\partial t_1}\frac{\partial X_j}{\partial t_1} \coth (X_i-X_j)\;,\label{EqG1}\\
\frac{\partial^2 X_i}{\partial t_1\partial t_2} &=& 2\sum_{j=1,j\neq i}^N\frac{\partial Xi}{\partial t_1}\frac{\partial X_j}{\partial t_2}\coth (X_i-X_j)\;.\label{EqG2}
\end{eqnarray}
The Lagrangians associated to these two equations of motion are
\\
\\
\textbf{\emph{HRS}}:
\begin{subequations}
\begin{eqnarray}\label{eqc2}
\mathscr L_{(t_1)}&=&\sum\limits_{i=1}^N\frac{\partial X_i}{\partial t_1}\ln\left| \frac{\partial X_i}{\partial t_1}\right|-\sum\limits_{i \ne j}^N\frac{\partial X_j}{\partial t_1} \ln\left | \frac{\sinh (X_i-X_j-\lambda)}{\sinh (X_i-X_j)}\right |,\;
\end{eqnarray}
\begin{eqnarray}\label{eqc3}
\mathscr L_{(t_2)}&=&\sum\limits_{i=1}^N\left(\frac{\partial X_i}{\partial t_2}\ln\left| \frac{\partial X_i}{\partial t_1}\right|
-\frac{1}{2\lambda }\left( \frac{\partial X_i}{\partial t_1}\right)^2+3\frac{\partial X_i}{\partial t_2}\right)\;\;\;\;\;\;\;\;\;\;\;\;\;\; \nn\\
&&-\sum\limits_{i \ne j}^N\left(\frac{\partial X_j}{\partial t_2}\ln\left | \frac{\sinh (X_i-X_j-\lambda)}{\sinh (X_i-X_j)}\right | + \frac{1}{2}\frac{\partial X_i}{\partial t_1}\frac{\partial X_j}{\partial t_1}\coth (X_i-X_j+\lambda)\right)\;.\nn\\
\end{eqnarray}
\end{subequations}
\textbf{\emph{HGF}}:
\begin{eqnarray}\label{eqc4}
\mathscr L_{(t_1)}&=&\sum\limits_{i=1}^N\frac{\partial X_i}{\partial t_1}\ln\left| \frac{\partial X_i}{\partial t_1}\right|+\sum\limits_{i \ne j}^N\frac{\partial X_j}{\partial t_1}\left( \ln \left|2\sinh (X_i-X_j)\right|\right),\;\;\;\;\;\;\;\;\;\;\;\;\;\;\;\;\;\;\;\;\;\;\;
\end{eqnarray}
\begin{eqnarray}\label{eqs5}
\mathscr L_{(t_2)}&=&\sum\limits_{i=1}^N\left(\frac{\partial X_i}{\partial t_2}\ln\left| \frac{\partial X_i}{\partial t_1}\right|
-\frac{1}{2}\frac{\partial X_i}{\partial t_2}\right) +\sum\limits_{i \ne j}^N\left(\frac{\partial X_j}{\partial t_2} \ln|2\sinh (X_i-X_j)|\right) \;.
\end{eqnarray}
\begin{acknowledgements}
This work is supported by Theoretical and Computational Science Center(TaCS), King Mongkut's University of Technology Thonburi(KMUTT), under Grant No. TaCS 2558-2.
\end{acknowledgements}

\end{document}